\documentclass[aps,prb,twocolumn,superscriptaddress,showpacs,floatfix]{revtex4-2}
\usepackage{amsmath,amssymb,amsfonts,amsthm}
\usepackage{graphicx}
\usepackage{bm}
\usepackage{bbm}
\usepackage{color}
\usepackage{mathtools}
\usepackage{dcolumn}   % needed for some tables
\usepackage{epstopdf}
\usepackage{bbold}
\usepackage[caption=false]{subfig}
\usepackage{xr}
\usepackage[colorlinks=true,linkcolor=blue,urlcolor=blue,citecolor=blue]{hyperref}
\definecolor{imcolor}{rgb}{0.5,0.,0.5}

\begin{document}

 \newcommand{\breite}{1.0} %  for twocolumn

\newtheorem{prop}{Proposition}
\newtheorem{cor}{Corollary} 

\newcommand{\be}{\begin{equation}}
\newcommand{\ee}{\end{equation}}

\newcommand{\bea}{\begin{eqnarray}}
\newcommand{\eea}{\end{eqnarray}}
\newcommand{\lt}{<}
\newcommand{\gt}{>} 

\newcommand{\Reals}{\mathbb{R}}     % Reals
\newcommand{\Com}{\mathbb{C}}       % Complex #
\newcommand{\Nat}{\mathbb{N}}       % Natural #

\newcommand{\id}{\mathbboldsymbol{1}}    

\newcommand{\Real}{\mathop{\mathrm{Re}}}
\newcommand{\Imag}{\mathop{\mathrm{Im}}}

\def\O{\mbox{$\mathcal{O}$}}   % Order epsilon ... 
\def\F{\mathcal{F}}			% FourierTrafo
\def\sgn{\text{sgn}}

\newcommand{\deo}{\ensuremath{\Delta_0}}
\newcommand{\dea}{\ensuremath{\Delta}}
\newcommand{\ak}{\ensuremath{a_k}}
\newcommand{\ad}{\ensuremath{a^{\dagger}_{-k}}}
\newcommand{\sx}{\ensuremath{\sigma_x}}
\newcommand{\sz}{\ensuremath{\sigma_z}}
\newcommand{\spl}{\ensuremath{\sigma_{+}}}
\newcommand{\smi}{\ensuremath{\sigma_{-}}}
\newcommand{\alk}{\ensuremath{\alpha_{k}}}
\newcommand{\bk}{\ensuremath{\beta_{k}}}
\newcommand{\ok}{\ensuremath{\omega_{k}}}
\newcommand{\vd}{\ensuremath{V^{\dagger}_1}}
\newcommand{\vi}{\ensuremath{V_1}}
\newcommand{\vo}{\ensuremath{V_o}}
\newcommand{\zc}{\ensuremath{\frac{E_z}{E}}}
\newcommand{\xc}{\ensuremath{\frac{\Delta}{E}}}
\newcommand{\xd}{\ensuremath{X^{\dagger}}}
\newcommand{\aok}{\ensuremath{\frac{\alk}{\ok}}}
\newcommand{\tpw}{\ensuremath{e^{i \ok s }}}
\newcommand{\tpe}{\ensuremath{e^{2iE s }}}
\newcommand{\tmw}{\ensuremath{e^{-i \ok s }}}
\newcommand{\tme}{\ensuremath{e^{-2iE s }}}
\newcommand{\epls}{\ensuremath{e^{F(s)}}}
\newcommand{\emis}{\ensuremath{e^{-F(s)}}}
\newcommand{\epl}{\ensuremath{e^{F(0)}}}
\newcommand{\emi}{\ensuremath{e^{F(0)}}}

\newcommand{\lr}[1]{\left( #1 \right)}
\newcommand{\lrs}[1]{\left( #1 \right)^2}
\newcommand{\lrb}[1]{\left< #1\right>}
\newcommand{\nbt}{\ensuremath{\lr{ \lr{n_k + 1} \tmw + n_k \tpw  }}}
\newcommand{\norm}[1]{\left|| #1 \right||}

\newcommand{\om}{\ensuremath{\omega}}
\newcommand{\dw}{\ensuremath{\Delta_0}}
\newcommand{\wbp}{\ensuremath{\omega_0}}
\newcommand{\dv}{\ensuremath{\Delta_0}}
\newcommand{\vbp}{\ensuremath{\nu_0}}
\newcommand{\vplus}{\ensuremath{\nu_{+}}}
\newcommand{\vminus}{\ensuremath{\nu_{-}}}
\newcommand{\wplus}{\ensuremath{\omega_{+}}}
\newcommand{\wminus}{\ensuremath{\omega_{-}}}
\newcommand{\uv}[1]{\ensuremath{\mathbf{\hat{#1}}}} % for unit vector
\newcommand{\abs}[1]{\left| #1 \right|} % for absolute value
\newcommand{\avg}[1]{\left< #1 \right>} % for average
\let\underdot=\d % rename builtin command \d{} to \underdot{}
\renewcommand{\d}[2]{\frac{d #1}{d #2}} % for derivatives
\newcommand{\dd}[2]{\frac{d^2 #1}{d #2^2}} % for double derivatives
\newcommand{\pd}[2]{\frac{\partial #1}{\partial #2}} 
% for partial derivatives
\newcommand{\pdd}[2]{\frac{\partial^2 #1}{\partial #2^2}} 
% for double partial derivatives
\newcommand{\pdc}[3]{\left( \frac{\partial #1}{\partial #2}
 \right)_{#3}} % for thermodynamic partial derivatives
\newcommand{\ket}[1]{\left| #1 \right>} % for Dirac bras
\newcommand{\bra}[1]{\left< #1 \right|} % for Dirac kets
\newcommand{\braket}[2]{\left< #1 \vphantom{#2} \right|
 \left. #2 \vphantom{#1} \right>} % for Dirac brackets
\newcommand{\matrixel}[3]{\left< #1 \vphantom{#2#3} \right|
 #2 \left| #3 \vphantom{#1#2} \right>} % for Dirac matrix elements
\newcommand{\grad}[1]{{\nabla} {#1}} % for gradient
\let\divsymb=\div % rename builtin command \div to \divsymb
\renewcommand{\div}[1]{{\nabla} \cdot \boldsymbol{#1}} % for divergence
\newcommand{\curl}[1]{{\nabla} \times \boldsymbol{#1}} % for curl
\newcommand{\laplace}[1]{\nabla^2 \boldsymbol{#1}}
\newcommand{\vs}[1]{\boldsymbol{#1}}
\let\baraccent=\= % rename builtin command \= to \baraccent
%%%%%%%%%%%%%%%%%%%%%%%%%%%%%%%%%%%%%%%%%%%%%
% End Definitions
%%%%%%%%%%%%%%%%%%%%%%%%%%%%%%%%%%%%%%%%%%%%%

%\newcommand{\IM}[1]{\texttt{\textcolor{Green} #1}}
%\newcommand{\KA}[1]{\texttt{\textcolor{RoyalBlue} #1}}
\def\red#1{{\textcolor{red}{#1}}}
%\newcommand{\IMcomment}[1]{\texttt{\color{} #1}}

%Title of paper
\title{Effect of quasiperiodic and random noise on many-body dynamical decoupling protocols}% Force line breaks with \\

\author{Tristan Martin}
 \email{tristan.martin@umontreal.ca}
 \affiliation{Physics Department, McGill University, Montreal, QC, Canada H3A 2T8}
\author{Ivar Martin}
 \affiliation{Material Science Division, Argonne National Laboratory, Argonne, IL 08540, USA}
\author{Kartiek Agarwal}
 \affiliation{Physics Department, McGill University, Montreal, QC, Canada H3A 2T8}

\date{\today}

\begin{abstract}
Symmetries (and their spontaneous rupturing) can be used to protect and engender novel quantum phases and lead to interesting collective phenomena. In Ref.~\cite{agarwal2020dynamical}, the authors described a general dynamical decoupling (polyfractal) protocol that can be used to engineer multiple discrete symmetries in many-body systems. The present work expands on the former by studying the effect of quasiperiodic and random noise on such a dynamical scheme. We find generally that relaxation of engineered symmetry generators proceeds by i) an initial relaxation on microscopic timescales to a prethermal plateau whose height is independent of noise, ii) a linear relaxation regime with a noise-dependent rate, followed by iii) a slow logarithmic relaxation regime that is only present for quasiperiodic noise. We glean the essential features of these regimes via scaling collapses and show that they can be generally explained by the spectral properties of the various noise waveforms considered. In particular, the quasiperiodic noise is characterised by highly time dependent spectrum with a noise floor that mimics white noise, and peaks that grow sharper with time. We argue that both the noise floor and peaks contribute to the initial linear-in-time relaxation while the logarithmic regime is initiated when the peaks become sufficiently well resolved and cease to contribute to further relaxation. We provide numerical evidence to justify these findings. 
\end{abstract}

\maketitle

\section{Introduction}

Understanding the phase structure of quantum systems in non-equilibrium settings is an important theoretical goal. Contrary to prior expectations, we now know that driven many-body quantum systems can avoid the fate of rapidly heating up to infinite temperature~\cite{moessner2017equilibration}, and over an intermediate but exponentially long prethermalization window~\cite{bukov2015universal,mori2016rigorous,kuwahara2016floquet,abanin2017rigorous}, be described by an effective Floquet Unitary and/or Hamiltonian, which in turn can espouse their own rich phase diagram~\cite{khemani2016phase,else2017prethermal,else2016floquet,yao2017discrete,ponte2015many}. This has been made possible, in large part, due to the theoretical discoveries of many-body localization~\cite{Basko,PalHuse,oganesyanhuselocalization}, and prethermal behavior in systems driven at high frequencies~\cite{kuwahara2016floquet,abanin2017rigorous}. A prominent recent example of such phenomena is the discrete time-translational symmetry breaking in time crystals~\cite{khemani2016phase,else2016floquet}, which has also been observed experimentally~\cite{zhang2017observation,choi2017observation}. 
% and the experimentally discovered photo-induced superconductivity; see Refs.~\cite{} for more examples. 

Much like their equilibrium counterparts, the richness of the non-equilibrium phases realized can be characterized by the symmetries of the Floquet system. In certain cases, the symmetry of the parent Hamiltonian is rich enough such that the role of driving is to simply push the system towards exhibiting the pheonomon of interest---this occurs for instance, in superconducting materials where driving optical phonons can lead to high temperature superconductivity~\cite{fausti2011light}, or for instance, in the case of topological wires supporting a fermion parity symmetry, which can be nudged into the Kitaev phase to support Majorana zero modes~\cite{agarwaldoublebraiding}.
On the other hand, driving can also be used to realize a greater set of symmetries than the parent Hamiltonian to achieve sought after phenomena. This was illustrated, for instance, in Ref.~\cite{friedman2020topological}
which showed how spin chains with $\mathcal{Z}_2 \times \mathcal{Z}_2$ symmetry weakly broken by disorder can stabilize topological spin-$1/2$ edge modes by restoring the symmetry using quasiperiodic driving. 

%Recently, two of the authors proposed a dynamical decoupling scheme, dubbed the polyfractal protocol, which helps realize multiple discrete symmetries by driving. The protocol was shown 

A natural question that arises is the role of noise and whether and how it may destabilize such desired dynamical behavior. In this work, we examine this question carefully in the context of a dynamical decoupling protocol proposed by two of the present authors and dubbed the polyfractal protocol. In particular, Ref.~\cite{agarwal2020dynamical} proposed this protocol to elevate a given set of discrete unitaries $X_i$ satisfying $X^2_i = \mathbb{1}$ to the role of symmetry generators of an effective Floquet many-body Hamiltonian. This can then be leveraged to support multiple Majorana zero modes in a fermionic one-dimensional system protected by additional symmetries~\cite{agarwaldoublebraiding},
for instance. Here we examine the efficacy of such driving in the presence of noise and how prethermalization pheonomena, and the associated symmetries, are affected. 

Specifically, the polyfractal protocol involves repeatedly driving the many-body system by periodic application of a set of $n_s$ unitaries $X_i$ at times $2^{(j-1)n_s + i-1} T_0$, where $T_0 \ll 1/\norm{h}$ is a driving period shorter than the inverse local energy scale of the physical system, given by $\norm{h}$, and $j = 1,..,n_f$ with $n_f$ being the number of `fractal layers' or the number of times each unitary $X_i$ is applied over the course of the entire Floquet period $T_f = 2^{n_s n_f} T_0$. The protocol renders $X_i$ as effective symmetry generators of the Floquet Hamiltonian $H_F$, albeit rotated by an angle $\sim \mathcal{O} \left( T^{n_f+1} \right)$. The protocol works in a rather intuitive manner analogous to the usual dynamical decoupling schemes~\cite{slichter2013principles,khodjasteh2005fault,khodjasteh2010arbitrarily}---
the application of $X_i$ repeatedly flips the signs of terms in the physical Hamiltonian that are odd under $X_i$, which are subsequently suppressed as a consequence of averaging over the Floquet period. Similar protocols have also been used to engineer global rotational~\cite{waugh1968approach}
and gauge symmetries~\cite{kasper2020non}. In this work, we examine the role of imperfect driving in this system, specifically by allowing for the times at which the unitaries $X_i$ are applied to deviate randomly from the prescribed times. 

In this work we particularly focus on quasiperiodic noise. This kind of noise has been previously studied in the context of many-body localized systems where it is seen to result in slow relaxation of operators~\cite{dumitrescu2018logarithmically,long2021many}, to effect multiple dynamical symmetries over a prethermalization window~\cite{else2020long}, to stabilize multiple topological edge modes~\cite{peng2018time,dumitrescu2021realizing}, and to realize multiple effective spatial dimensions~\cite{martin2017topological,agarwal2017localization}. Specifically, we consider a version of such noise that can be generated using Fibonacci recursion relations and allows for efficient simulation of relaxational dynamics of operators over exponentially long timescales. In particular, we assume that $X_i$ are applied after long and short intervals of length $T_0 (1 + \epsilon/\phi)$ and $T_0 (1 - \epsilon)$, alternating in the characteristic Fibonacci sequence~\cite{levine1984quasicrystals,levine1986quasicrystals,socolar1986quasicrystals} (to be prescribed in more detail below), with $\epsilon$ being a small parameter that can be used to smoothly control the amplitude of such noise. The precise extent of the long and short durations ensure that the average application time of $X_i$ occurs at times $t = nT_0$ after $n$ applications of the driving unitaries. For comparison, we also consider synchronous and asynchronous noise. These involve randomizing the time of application of $X_i$ by a Gaussian of width $\epsilon T_0$. In the asynchronous case, there is no perfect clock with period $T_0$ that tries to calibrate future applications of $X_i$ and deviations from the perfect application at times $t = n T_0$ grow indefinitely. In the synchronous case, a perfect clock is used to make sure that deviation from prescribed application $t = nT_0$ is always $\sim \mathcal{O} \left( \epsilon T_0 \right)$. 

To understand the effect of the noise we simulate the relaxation of the unitaries $X_i$ which, in the ideal decoupling limit, should serve as perfect integrals of motion, at least over an exponentially long prethermal window. In particular, we find the following three regimes that describe the relaxation of these effective symmetry generators---i) an initial rapid relaxation that occurs over microscopic time scales accompanied by a decay $\sim \mathcal{O} \left(T^{n_f + 1}\right)$ of $X_i$ in line with the findings of Ref.~\cite{agarwal2020dynamical}, ii) an intermediate plateau which gives way to a linear in time relaxation that occurs at a rate $\sim \mathcal{O} \left( \epsilon^2 T^2_0\right)$, followed by iii) a final slow logarithmic relaxation regime which occurs only for quasiperiodic driving and which kicks in at time $\sim \mathcal{O} \left(1/\epsilon T_0\right)$. This last regime appears to continue until the total relaxation of $X_i$, if it occurs at all. A summary of these relaxation regimes and our main numerical findings can be seen in Fig.~\ref{fig:caricature}

Although we simulate many-body systems driven by unitaries $X_i$ which involve operators spanning the extent of the system, the dynamics can easily be described equivalently in terms of a local drive that acts effectively by periodically modulating the sign of local terms in the physical Hamiltonian that are odd under $X_i$. We find that the spectral properties of this noise can, in fact, explain the entire relaxation curve outlined above. In this sense, our work sheds light on the key features that distinguish random noise from quasiperiodic noise, besides providing a simple physical understanding of relaxation processes in these dynamical systems, and the robustness of these protocols to noise, which is relevant for practical implementation of such ideas. 

This manuscript is organized as follows. In Sec.~\ref{sec:reviewpolyfractal}, we first review the polyfractal protocol and the intuition behind it. We also briefly discuss the connection to time crystals for $n_s = 1$, and $X_1$ is the global spin flip operator. In Sec.~\ref{sec:noisespectrum}, we discuss the spectral properties of the types of noise under examination, particularly elaborating on the properties of the quasiperiodic noise and how numerical simulations are performed. In Sec.~\ref{sec:numerics}, we discuss numerical results on simulating a generic spin chain Hamiltonians driven by one or many $X_i$, and over multiple fractal layers. We elucidate the complete relaxation curve by using appropriate scaling collapses and also explain the results obtained by alluding to the spectral properties of the noise studied. We also show that even single spins when driven analogously display the behavior we find in the many-body spin chain setting. In Sec.~\ref{sec:conclusions}, we conclude with a discussion of the results and their implications, and outline certain other findings that motivate further investigation.

\section{Review of polyfractal protocol}
\label{sec:reviewpolyfractal}

The polyfractal protocol corresponds to the time evolution of the system under its physical Hamiltonian $H$, but punctuated by the application of unitary operations $X_i$ periodically, at increasingly longer periods. To begin with, we can understand the effect of this protocol by considering the application of a single unitary $X_1$ periodically at time $T_0$. The Floquet unitary describing time evolution over the period $T_F = 2T_0$ is then given by 

\begin{align}
    U_F &= X_1 e^{-i H T_0} X_1 e^{-i H T_0} \nonumber \\ 
    &= e^{-i (A-B) T_0} e^{-i (A+B) T_0} \nonumber \\
    &= e^{-i A T_f + B' \mathcal{O} \left( T_0^2 \right) }.
    \label{eq:dyndecoupling}
\end{align}

Here, we expanded $H = A+B$ in terms of local terms that are even $(A)$ and odd $(B)$ under $X_1$, specifically, $X_1 A X_1 = A$, $X_1 B X_1 = -B$. The Baker-Campbell-Hausdorff (BCH) expansion can then be used to justify, in the limit of small $T$, that the Floquet unitary can be described by a Floquet Hamiltonian $H_F \approx A$ which is symmetric under $X_1$, while antisymmetric corrections $B'$ appear at order $\mathcal{O} \left( T^2_0 \right)$. We note that the fact that a (quasi-)local Floquet Hamiltonian can be obtained from the BCH expansion for a many-body system relies on the notion of a prethermalization window of duration~\cite{kuwahara2016floquet,abanin2017rigorous} $\sim e^{\norm{h}/T} $ over which $H_F$ is approximately conserved. Here $\norm{h}$ is the operator norm of the local operator that makes up the physical Hamiltonian after translations by multiples of the unit cell. 

The polyfractal protocol now corresponds to supplementing the above with with additional driving unitaries $X_2, X_3, ...$ applied at periods $2T_0, 4T_0, ...$ (`poly' in polyfractal) and further repeating all these $n_s$ unitaries at periods $2^{n_s+1} T_0, 2^{n_s+2} T_0, ...$ (`fractal' in polyfractal). For concreteness, the protocol involves applying $n_s$ unitaries $X_i$ at periods $T_i = 2^{(j-1)n_s + i-1} T_0$, where $i \in \left\{ 1, ..., n_s \right\}$, $j \in \left\{1,..., n_f \right\}$, and $n_f$ denotes the number of fractal layers. 

\begin{figure}[htp]
\includegraphics[width=2in]{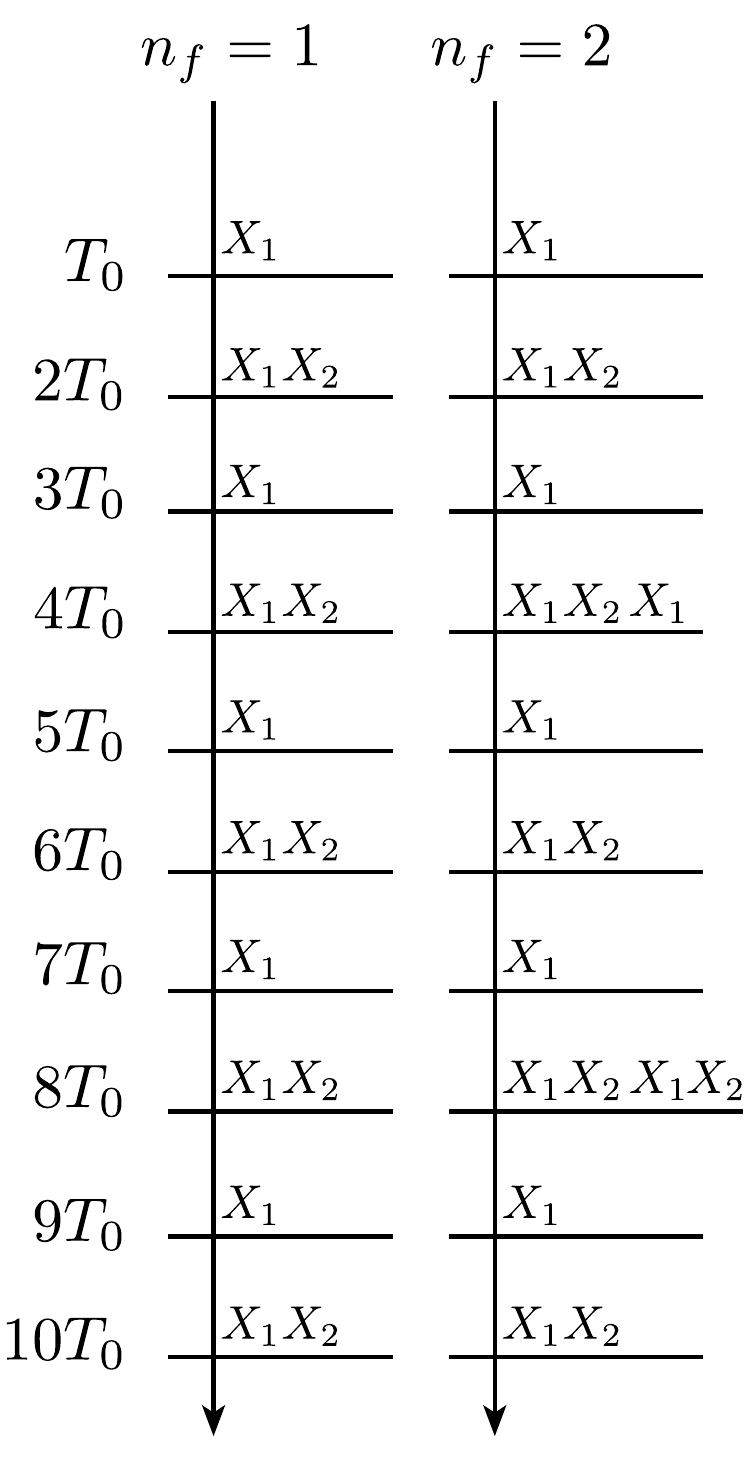}
\caption{Illustration of the polyfractal protocol used for dynamical decoupling. The protocol involves unitary evolution under the system's physical Hamiltonian, punctuated by the application of operators $X_i$ at specific times. The above illustrates when these operators are applied for the case where the number of operators, $n_s = 2$, and number of fractal layers, $n_f = 1,2$ up to $t = 10 T_0$. (Note that the period for the protocol at $n_f = 2$ is $16 T_0$; thus the entire period is not illustrated.)  Note that at times where the same operator $X_i$ appears an even number of times, it can be ignored by virtue of the fact that $X_i^2 = \mathbb{1}$.}
\label{fig:polyfrac}
\end{figure}

A particular illustration of the times at which these operators are applied is exhibited in Fig.~\ref{fig:polyfrac} for $n_s = 2$, but different fractal layers $n_f = 1,2$. Concretely, for $n_f = 1$, the operator $X_1$ is applied with period $T_0$, and $X_2$ with period $2T_0$; Ref.~\cite{agarwal2020dynamical} shows that an effective Hamiltonian then describes the dynamics stroboscopically at periods $T_F = 4 T_0$, and importantly, commutes with $X_1, X_2$ up to terms of order $\mathcal{O} \left( (T_0 \norm{h})^2 \right)$. For $n_f = 2$, the operators $X_1, X_2$ are reapplied, now at periods $4T_0, 8T_0$, respectively. The effective Hamiltonian in this case, that describes dynamics stroboscopically at the Floquet period $T_F = 16 T_0$, commutes with both $X_1, X_2$ to order $\mathcal{O} \left( (T_0 \norm{h})^3 \right)$.

%The polyfractal protocol corresponds to time evolution of the system under its physical Hamiltonian $H$, but punctuated by the application of the operations $X_i$ periodically with increasingly-longer periods related by factors of $2$. We assume that this Hamiltonian is a sum of local terms, with the operator norm bounded by $\norm{h}$. 

%Note that we aim for the unique set of operators $X_i$, $i \in [1,n_s]$ to be symmetry generators of the effective Hamiltonian obtained from this dynamical scheme. Periodic application of a particular $X_i$ allows one to suppress terms in the effective Hamiltonian that do not commute with $X_i$. In the same spirit, by applying multiple $X_i$ at periods $T_i = 2^{(i-1)} T_0$, we can make sure that the resultant Hamiltonian commutes with all $X_i$. Next, the dynamical suppression of non-commuting terms can be further enhanced by repeating this entire sequence at twice the period, and this represents a second `fractal layer' of protocols. In total, we imagine applying the unitary $X_i$ at period $T = 2^{i -1 + j n_s} T_0$, where $i \in \{1,...,n_s\}$, $j \in \{0,1,...,n_f -1 \}$, and $n_s$ and $n_f$ denote the total number of distinct operators $X_i$, and the number of fractal layers, respectively. 

This process cannot be continued indefinitely because that involves driving the system at smaller and smaller frequencies, and eventually leads to uncontrolled heating. In particular, there exists an optimal number of fractal layers given by $n^*_f \sim \text{log} \left( \frac{1}{n_s} \text{log}_\text{2} \left( \frac{1}{T_0 \norm{h}} \right) \right)$. 
%\IM{we should give here a ref and also a caveat. Doesn't the protocol work better than this estimate seems to imply?} 
For this  choice of $n_f$, the heating rate  is stretched-exponentially long in $1/(T_0 \norm{h})$; thus heating may be ignored for a long time within this scheme, while the time up to which the effective Hamiltonian may be assumed to commute with $X_i$, scales as $t_* \sim \norm{h} \left( \frac{1}{T_0 \norm{h}} \right)^{n^*_f}$, which grows polynomially in the drive frequency, but with a power that itself grows as $T_0 \rightarrow 0$. We further note that the numerical data~\cite{agarwal2020dynamical} consistently show a much longer, possibly exponential timescale over which the operators $X_i$ continue to be symmetry generators when this scheme is implemented which only appears to become finite when $T_0$ approaches $1$.

For the purposes of this work, it will be useful to partition $H$ into terms which transform as odd/even operators under $X_i$:  
\begin{align}
&H = \sum_{\vs{\epsilon}} A_{\vs{\epsilon}} \; \text{where} \; \vs{\epsilon} = (\epsilon_1, ..., \epsilon_{n_s}), \epsilon_i \in \{ 0,1 \}, \nonumber \\
&X_j A_{\vs{\epsilon}} X_j = (-1)^{\epsilon_j} A_{\vs{\epsilon}}. 
\end{align}
%This decomposition is unique if $X_i$s commute or anti-commute with one another, which we assume. %The natural setup for such operations is ensembles of spins-1/2s or Majorana fermions, but one may also imagine such a setup for usual complex fermions and hardcore bosons. 

and represent the Floquet unitary in time-ordered notation as 
\begin{align}
&U (T_F \equiv 2^{n_f n_s} T_0) = \mathcal{T} \left\{ e^{-i \int^T_0 dt \; \sum_{\vs{\epsilon}} A_{\vs{\epsilon}} f_{\vs{\epsilon}} (t) } \right\}, \nonumber \\
&\text{where} \; f_{\vs{\epsilon}} (t) = \pm 1 \;  \; \text{and} \; \int_0^{2^{n_s}T_0} f_{\vs{\epsilon}} (t) = \delta_{\vs{0}, \vs{\epsilon}}
\label{eq:Hphys}
\end{align}

Here $A_{\vs{0}}$ comprises of only terms even under all $X_i$ and is thus time-independent. It thus effectively serves as the `physical' Hamiltonian of the system, alongside local drive terms $A_{\vs{\epsilon}}$ for $\vs{\epsilon} \neq \vs{0}$. Thus, one can reconsider the problem in terms of a local physical Hamiltonian driven globally with local drives with noise spectra

\begin{align}
    S_\epsilon (\omega) \equiv \abs{f_{\vs{\epsilon}} (\omega)}^2.
\end{align}

%\IM{How does $A_{\vs{\epsilon}}$ enter?}
We note that in this work, we explicitly only seek to decribe the dynamics at times $t = nT_F$, or multiples of the Floquet period $T_F$. It is worth noting that it has also been shown rigorously, that for the case of a single driving unitary $X_1$, and $n_f = 1$, the stroboscopic dynamics at half the period, $T_F/2 = T_0$ are captured by the Floquet unitary $U_F \approx X_1 e^{-i H_F T_0}$ over an exponentially long prethermalization window, and where $[H_F,X_1]$ is exponentially small, $\sim e^{-\norm{h}/T_0}$. The breaking of the spatial $\mathcal{Z}_2$ symmetry of $H_F$ then translates to a broken discrete time-translation symmetry and serves as the paradigmatic example of a discrete time crystal. 

In this work, we focus only on the dynamics over the doubled period $T_F$ and the physics espoused by the effective Floquet Hamiltonian $H_F$. Next, we discuss the spectral form of the noises considered as this largely capture the physics of the problem at hand.  

\section{Noise studied and associated spectral properties}
\label{sec:noisespectrum}

With the protocol specified, we can now discuss various ways of perturbing the protocol with random perturbations to introduce inaccuracies or noise. In this work, we primarily consider implementation noise of a particular kind---the unitaries $X_i$ that are used to implement the protocol are applied at irregular times, departuring from the prescribed period in the polyfractal protocol. This form of noise is more amenable to numerical simulations---continuous noise is considerably more challenging to numerically simulate for long enough times to capture the behavior of interest. 

It is worth noting that any continuous noise will fall into roughly three categories---i) low frequency noise with $\omega \ll 2\pi/T$ is benign, provided the physical operators that it couples to are driven. In particular, the noise can couple to operators that are odd/even under the action of various $X_i$s. The coupling to operators that are odd in at least one $X_i$ will be suppressed significantly at frequencies below $2\pi / T$ when the driving scheme is implemented. Coupling to operators even in all $X_i$s cannot be corrected for and will lead to dephasing; this can be an issue in many physical systems where $1/f$ noise leads to deleterious dephasing~\cite{paladino20141fnoise}, but where spin-echo protocols can be devised to strongly suppress its negative effect. We assume that operators coupling to such noise sources are driven by carefully chosen $X_i$. ii) High frequency noise with $\omega \gg 2\pi/T$ is largely benign as it cannot lead to rapid heating since we assume $T \lesssim 1/\norm{h}$ and thus such noise corresponds to frequencies that would require multi-spin rearrangements to be absorbed by the system; this is a higher order process in perturbation theory in $1/\norm{h}T_0$ and is thus inefficient at heating the system, and finally, iii) noise at intermediate frequencies $\omega \sim 2\pi/T$, which we attempt to capture within our approach (which we note also contain low frequency components). 

In particular, we model the role of noise by considering three kinds of irregularities in the implementation of $X_i$ which we describe below.  

%It is clear that the protocol is not changed with the addition of static perturbations to the Hamiltonian as long as this does not significantly alter the norm of the terms $\norm{h}$ or the locality of the Hamiltonian. Another form of noise that the system could encounter is the introduction of dynamical fluctuations in the physical Hamiltonian $H$. Unfortunately, such fluctuations would be remarkably challenging to thoroughly investigate numerically. In particular, 

\subsection{Asynchronous and synchronous noise}

\begin{figure}
    \centering
    \includegraphics[width=3.5in]{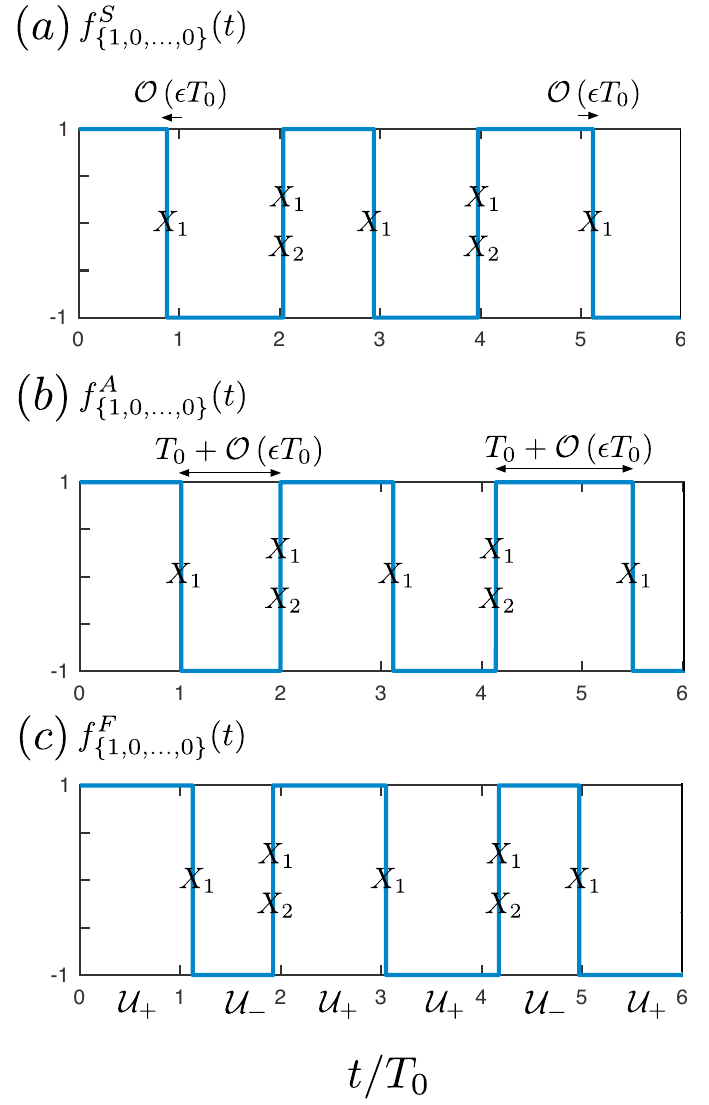}
    \caption{(a) Synchronous noise. All deviations from the perfect period are $\mathcal{O} \left( \epsilon T_0 \right)$. (b) Asynchronous noise. All intervals between successive applications of the driving unitaries are of length $T_0 + \mathcal{O} \left( \epsilon T_0 \right)$. (c) Fibonacci noise. All intervals are of length $T_+$ or $T_-$ (indicated below by the appropriate unitary). For all cases, we show the drive $f_{\vs{1,0,...,0}} (t)$ associated with the term odd in $X_1$ and even in all other driving unitaries. Here $T_0 = 1, \epsilon = 0.2$.}
    \label{fig:noiseform}
\end{figure}

We consider synchronous noise which involves deviations in the implementation times of $X_i$ from the prescribed multiples of period $T_0$ by $\mathcal{O} \left ( \epsilon T_0 \right)$. In particular, the deviation is described by a Gaussian random variable with distribution $N (0, \epsilon T_0) $ of zero mean and standard deviation $\epsilon T_0$. 
%This is the natural scenario when considering a realistic implementation of the protocol involving a timer phase locked to a laser, for instance. \IM{realistic implies that this has been done already, or smth very similar has been done. Has it? The description is a bit obscure} 
We illustrate such noise graphically by plotting the waveform corresponding to $f_{(1,0,...,0)} (t)$, the factor that modulates the term $A_{(1,0,...,0)} (t)$ which is only odd in $X_1$ and even in all other $X_i$ in the time-dependent Hamiltonian of Eq.~(\ref{eq:Hphys}). In the ideal case, this function oscillates between $\pm 1$ with period $T_0$. In the presence of synchronous noise, the waveform $f^S_{(1,0,...,0)} (t)$ is obtained which we plot in Fig.~\ref{fig:noiseform} (a). 

We also consider asynchronous noise wherein the temporal deviations from ideal application times accrue. Specifically, each subsequent application of the unitary $X_i$ is performed at a time that deviates from $T_0$ by a random Gaussian of standard deviation $\epsilon T_0$. Note that in both the synchronous and asynchronous cases, all operators $X_i$ that are meant to be applied in the polyfractal protocol simultaneously, continue to be applied simultaneously, but with the slight temporal deviation characterized by $\epsilon T_0$. We illustrate this noise in Fig.~\ref{fig:noiseform} (b) and the times at which operators $X_1,X_2$ are applied for the case $n_s = 2, n_f = 1$. 

\begin{figure}
    \centering
    \includegraphics[width=3in]{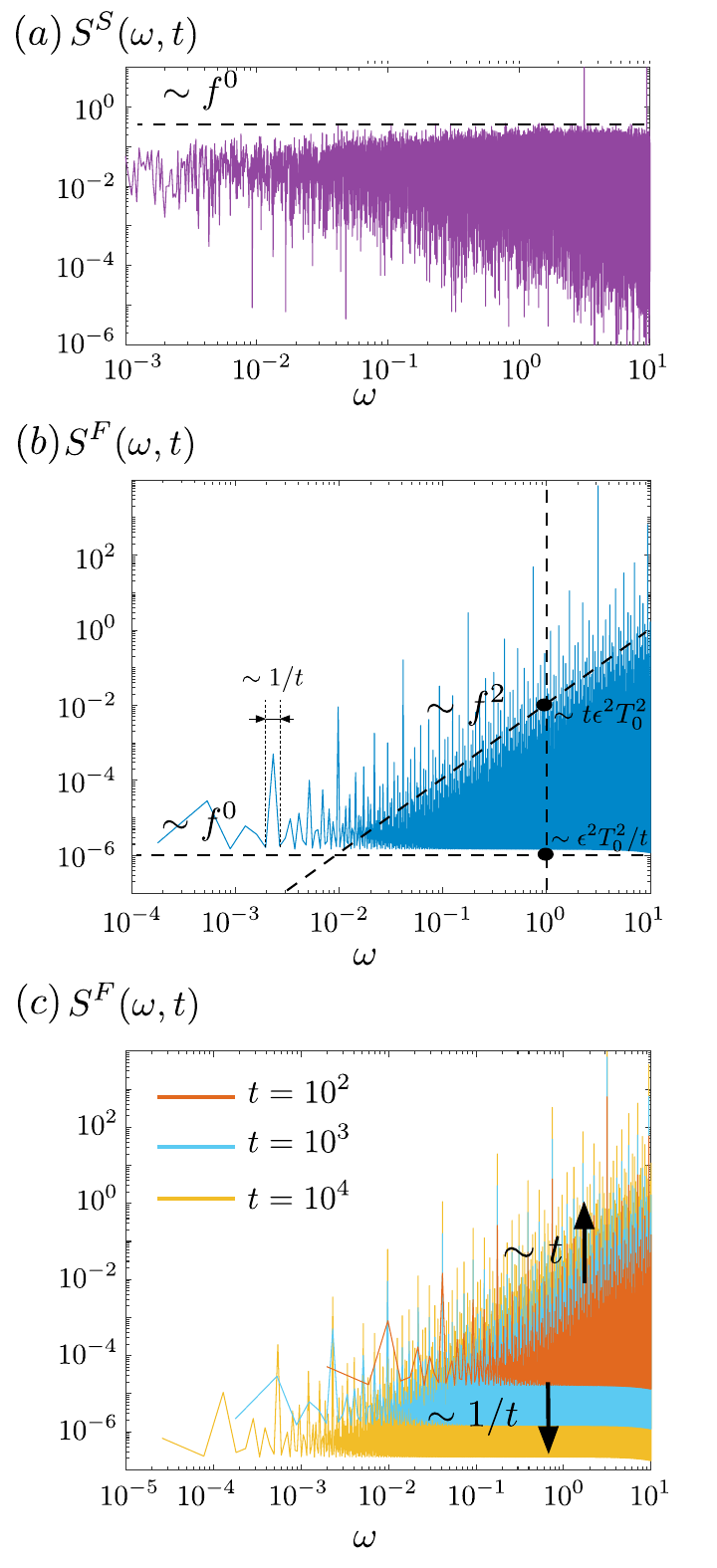}
    \caption{ The spectral function evaluated numerically, as function of the frequency $\omega$ and maximum time $t$ of the noise waveform $f_{\vs{1,0,...,0}} (t)$, associated with (a) synchronous noise and (b) Fibonacci noise. Here $T_0 = 1, \epsilon = 0.1, t = 10^3$. Both synchronous and asynchronous noises are featureless except for a peak near $\omega = \pi$ (here $T_0 = 1$) (we show only the first to avoid redundancy). Fibonacci noise shows a random noise floor which is flat in frequency, and a peak structure that grows as $\omega^2$. The spectral function changes considerably with time, $t$, with the floor decreasing as $1/t$ and the peaks increasing in amplitude as $t$; see (c). The peaks have a self-similar structure, and width $\sim 1/t$. The noise amplitude at $\mathcal{O} (1)$ frequencies is proportional to $\epsilon^2 T_0^2$ in all cases. The total spectral weight appears generally to be larger for random noise as opposed to Fibonacci noise; in this case by a factor of $\approx 2$.}
    \label{fig:noisespectrum}
\end{figure}

There is no difference in the form relaxation curve of the effective symmetry generators $X_i$ in the presence of these two noise forms, although for the same values of $\epsilon, T_0$, relaxation under asychronous noise usually appears to be a bit slower. This can be surmised as a consequence of the fact that the spectral properties of these two noise forms is the same---these are both examples of `white noise' with a flat frequency spectrum---with some difference in the actual amplitude of the spectrum. The frequency spectra of synchronous noise is shown in Fig.~\ref{fig:noisespectrum} (a).

\subsection{Fibonacci noise}

We now detail another form of implementation noise which is systematic but effectively appears to be random at short times; this kind of implementation noise forms the main thrust of our work. A major advantage of such a noise form is that it allows for much longer simulation times as the time-evolution matrix can be obtained by recursive matrix multiplications at multiples of $T_0$ times a Fibonacci number $F_n$, the latter of course growing exponentially with $n$. This form of noise has been used to study quasiperiodic driving in many-body quantum systems recently. 
%\IM{we should justify better, beyond the fact that it is easier to simulate. Should we say that we expect interesting physics? connection to time qcrystals? }

In particular, we consider `Fibonacci noise' wherein successive applications of unitaries $X_i$ are no longer periodic but separated by either a longer time interval  of duration $T_+ = T_0 ( 1 + \epsilon/\phi)$ or a shorter interval of duration $T_- = T_0 ( 1- \epsilon)$, which appear in the characteristic Fibonacci recursion pattern. Here $\phi$ is the golden mean, and the form of these durations is chosen such that application of the unitaries $X_i$ at long time continues to have the period $T_0$, on average. (Note that setting a different ratio for $T_+/T_-$ by using another constant instead of $\phi$ would lead to a renormalized $T_0$ and $\epsilon$.)
%Defining the time-evolution operators $\mathcal{U}_\pm = e^{-iHT_\pm}$ 
%

To better understand this noise and how it can be numerically simulated, we consider some examples. Let us first consider the Fibonacci distortion of the case of driving with a single unitary $X_1$, applied at period $T_0$ (the case of $n_s = 1, n_f = 1$). Defining the time-evolution operators $\mathcal{U}_\pm = e^{-iHT_\pm}$ for the short and long time steps under the physical Hamiltonian, and $U_1 = X_1 \mathcal{U}_-$, $U_2 = X_1 \mathcal{U}_+$, the time evolution for the first few Fibonacci recursions is given by

\begin{align}
U_3 &=\underset{U_1}{ \underbrace{X_1 \mathcal{U}_{-}}} \underset{U_2}{\underbrace{X_1 \mathcal{U}_{+}}} \nonumber \\
    U_4 &= \underset{U_2}{ \underbrace{X_1 \mathcal{U}_{+}}} \underset{U_3}{\underbrace{X_1 \mathcal{U}_{-} X_1 \mathcal{U}_{+}}} \nonumber \\
    U_5 &= \underset{U_3}{ \underbrace{X_1 \mathcal{U}_{-}X_1 \mathcal{U}_{+}}} \underset{U_4}{\underbrace{X_1 \mathcal{U}_{+} X_1 \mathcal{U}_{-} X_1 \mathcal{U}_{+}}} 
    \label{eq:fibns1}
\end{align}

This corresponds to the time evolution unitary given simply by $U_n = U_{n-2} U_{n-1}$ for  $n>2$. Clearly, long and short periods of time evolution under the physical Hamiltonian appear in a Fibonacci recursion pattern, with the operator $X_1$ being applied after each such period. In this case, the recursion relation is particularly simple and can be used to evaluate the time-evolution unitary up to exponentially long times.

It is also instructive to consider the Fibonacci distortion of the case for $n_s =2$, $n_f = 1$ (in the undistorted case, the system is driven periodically by unitaries $X_1, X_2$ applied with periods of $T_0, 2T_0$, respectively). In this case, the time evolution for the first few Fibonacci recursions is given by

\begin{align}
    U_3 & =\underset{U_1^* \neq U_1}{ \underbrace{X_2 X_1 \mathcal{U}_{-}}} \underset{U_2}{\underbrace{X_1 \mathcal{U}_{+}}} \nonumber \\
    U_4 & = \underset{U_2^* = U_2}{ \underbrace{X_1 \mathcal{U}_{+}}} \underset{U_3}{\underbrace{X_2 X_1 \mathcal{U}_{-} X_1 \mathcal{U}_{+}}} \nonumber \\
    U_5 & = \underset{U_3^* \neq U_3}{ \underbrace{X_1 \mathcal{U}_{-}X_2 X_1 \mathcal{U}_{+}}} \underset{U_4}{\underbrace{X_1 \mathcal{U}_{+} X_2 X_1 \mathcal{U}_{-} X_1 \mathcal{U}_{+}}}
    \label{eq:fibns2}
\end{align}

%\IM{need to define $U^*$} 
Here we note that the recursion relation becomes more complex. Again, long and short periods of time evolution follow in the same pattern in this case as for $n_s = 1, n_f = 1$ [compare Eqs.~(\ref{eq:fibns1}) and (\ref{eq:fibns2})] but in this case, these periods are followed by applications of the unitaries $X_1$ after each period of evolution by the physical Hamiltonian, \emph{and} $X_2$ after each second such period of evolution. 
%\IM{need to rephrase somehow}. 
In this case, a modified recursion relation $U_n = U^*_{n-2} U_{n-1}$ applies where $U^*_{n-2}$ can be either one of $U^A_n, U^B_n$ defined at each recursion step and which differ from one another by the placement of operators $X_1, X_2$ but not in the order of $\mathcal{U}_\pm$. (One can understand the presence of two such unitaries by noting that $X_1$ appears after every period of evolution by $\mathcal{U}_\pm$, but $X_2$ may appear at odd or even steps in these two copies.) 

The time evolution can be performed similarly for larger $n_s, n_f$ cases. Note that the general structure is the same---the order of long and short periods of time evolution remains the same for all cases and is given by the Fibonacci recursion pattern, and after each $2^{i-1}$ periods of evolution (by either $\mathcal{U}_\pm$), unitaries $X_i$ are applied. The lack of a simple recursion relation makes the simulation for larger $n_s, n_f$ considerably more challenging but a general prescription can be formulated. More details can be found in Apps.~\ref{app:ns1} for $n_s = 1$ and~\ref{app:ns2} for $n_s = 2$, respectively. We also plot the function $f^F_{(1,0,...,0)} (t)$ for the Fibonacci case in Fig.~\ref{fig:noiseform} (c) for illustration. 

\subsection{Noise Spectrum}

We now examine the spectral properties of the noise forms introduced. In particular, both synchronous and asynchronous noises correspond to a largely featureless white noise spectrum, with an amplitude $\sim \epsilon^2 T_0$. We plot the former in Fig.~\ref{fig:noisespectrum} (a). 
%\IM{wasnt synchronous a much weaker noise?? woudl be good to plot both, not just say in words} 

The Fibonacci noise spectrum is more interesting. The Fibonacci ``noise" is not random at all and thus its spectrum can depend strongly on the total time $t$ over which the waveform is generated. In particular, at all times, the spectral function $S^F(\omega,t)$ has a noise floor which is approximately independent of the frequency. In addition, there are peaks with an amplitude $\sim \epsilon^2 T_0^2 \omega^2 t$ and a width $\sim 1/t$ related to the total duration of the waveform $t$. The peaks show a self-similar structure with the most prominent peaks appearing at frequencies $\approx \pi/(T_0 \phi^n)$. It is evident that on a logarithmic scale, these peaks appear equally spaced. Thus the density of peaks, $N(\omega) \sim 1/\omega$ and the typical amplitude of the noise spectrum can be approximated by $N(\omega) \cdot 1/t \cdot \epsilon^2 T_0^2 \omega^2 t = \epsilon^2 T_0^2 \omega$, and is independent of time. The noise floor has an amplitude which can be surmised from the amplitude of the peaks at the lowest frequencies, $\omega \sim 1/t$, and is frequency independent, thus mimicking white noise with an amplitude that \emph{decreases in time} as $\sim 1/t$. 

%The Fibonacci noise spectrum is more interesting. It is important to note that Fibonacci ``noise" is not random at all and thus its spectrum can depend strongly on the total time $t$ over which the waveform is generated. In particular, at all times, the spectral function $S^F(\omega,t)$ has a noise floor which is approximately independent of the frequency. In addition, there are peaks that grow as $\sim \omega^2$ with a width $\sim 1/t$ related to the total duration of the waveform $t$. The peaks show a self-similar structure with the most prominent peaks appearing at frequencies $\approx \pi/(T_0 \phi^n)$. Crucially, as time $t$ increases, more peaks get resolved and effectively absorb the spectral weight from the noise floor, in stark contrast to random white noise. As a result, the peaks increase in amplitude as $\sim t$ while the noise floor decreases concomitantly as $\sim 1/t$ while the total spectral density to remains invariant in time, as expected. Finally, all noise forms show an amplitude that scales as $\sim \epsilon^2 T_0^2$ at frequency $\omega \sim \mathcal{O} (1)$. 

Note that all results have been shown for the spectral function associated with the modulation $f_{\vs{\epsilon} = (1,0,...,0)} (t)$ which is associated with the term in the Hamiltonian that is odd in $X_1$ and even in all other $X_i$. However, the general features of the results presented hold true for all $f_{\vs{\epsilon}} (t)$ at $\mathcal{O} (1) \ll 2\pi/T_0 $ frequencies---this can be noted from the fact that all $f_{\vs{\epsilon}} (t)$ differ from one another by sign changes that occur on the timescale $T_0$, while the essential information of the Fibonacci modulation is evident only due to correlations over much longer times. Thus, for the purposes of frequencies that particularly affect the relaxation dynamics in the system, all functions $f_{\vs{\epsilon}} (t)$ have spectral properties that scale similarly. We will see that these features explain much of our numerical findings on the relaxation of the driving unitaries (and effective symmetry generators) $X_i$. 

\section{Numerics on spin chains}
\label{sec:numerics}

We now implement the protocols along with noise forms outlined above on a many-body system, a spin-$1/2$ chain of length $L$. The spin chain Hamiltonian we study is composed only of local terms, which nevertheless are more easy to describe in terms of Majorana operators $\gamma_i$ which satisfy the algebra $\left \{ \gamma_a, \gamma_b \right \} = 2 \delta_{ab}$. Noting the Jordan-Wigner transformation
\begin{align}
\gamma_{2j-1} &\equiv a_j = \left ( \prod_{k = 1}^{j - 1} \sigma_{k}^z \right ) \sigma_j^x, \\ 
\gamma_{2j} &\equiv b_j = \left ( \prod_{k = 1}^{j - 1} \sigma_{k}^z \right ) \sigma_j^y,
\end{align}
the Hamiltonian is explicitly given by
\begin{align}
\mathcal{H} = \sum_{j,k \leq 4} e^{-k+1} \left (-i\gamma_j \gamma_{j+k} \right ) + V \gamma_j \gamma_{j+1} \gamma_{j+2} \gamma_{j+3},
\end{align}
where we do not allow for terms odd in the number of Majorana operators. This gives the Hamiltonian a global fermion Parity symmetry $[H,P_z] = 0$ with 
\begin{equation}
P_Z = \prod_{j = 1}^{L} \sigma_j^z = (-i)^L \prod_{j = 1}^{2L} \gamma_j = (-i)^L \prod_{j =1}^{L} a_j b_j.
\end{equation}
This  ensures that the Hamiltonian in the spin language is local as well, and is composed of a whole host of terms with range up to next nearest-neighbor couplings. Without loss of generality, we work in the parity sector $P_z = 1$. 
The system is driven with the following unitary operators
\begin{align}
X_1 = P_{X} & = \prod_{j = 1}^{L} \sigma_j^x = (-i)^{L/2} \prod_{j = 1}^{L/2} b_j a_{j+1} \\
X_2 = P_{Z2} & = \prod_{j = 1}^{L/2} \sigma_{2j}^z = (-i)^{L/2} \prod_{j = 1}^{L/2} a_{2j} b_{2j}
\end{align}

Note that $[X_1,X_2] = 0$ for $L = 4n$ and $\left \{ X_1,X_2 \right \} = 0$ for $L = 4n+2$. 

\begin{figure}
    \centering
    \includegraphics[width=3in]{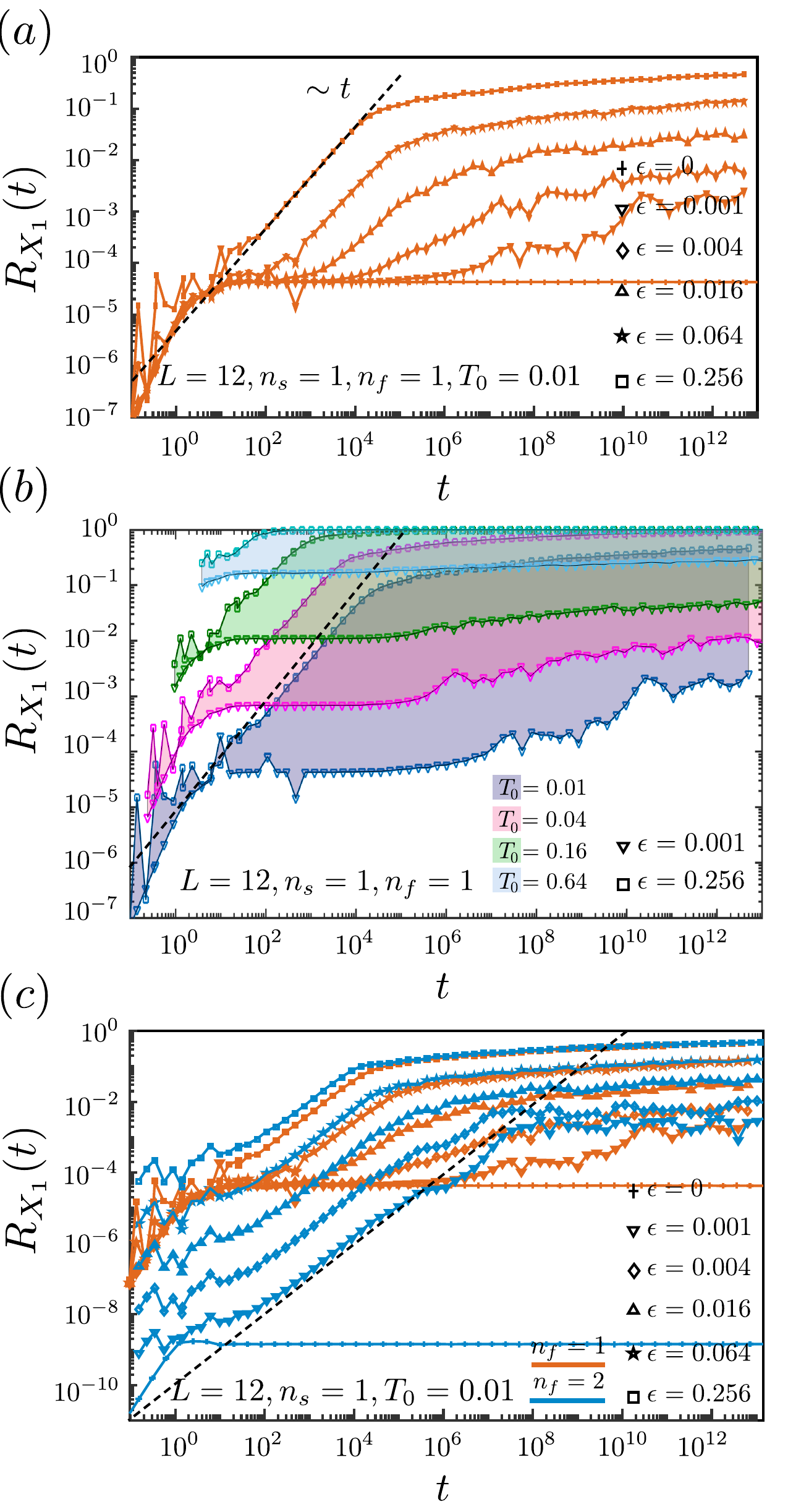}
    \caption{Relaxation dynamics of $X_1$ when the system is driven by the single unitary, that is, $n_s = 1$. Black dashed lines indicate $\sim t, t^2$ decay. All relevant parameters are indicated in the plots; see the main text for further discussion.}
    \label{fig:ns1_unscaled}
\end{figure}

\subsection{General Observations}

\begin{figure}
    \centering
    \includegraphics[width=3.4in]{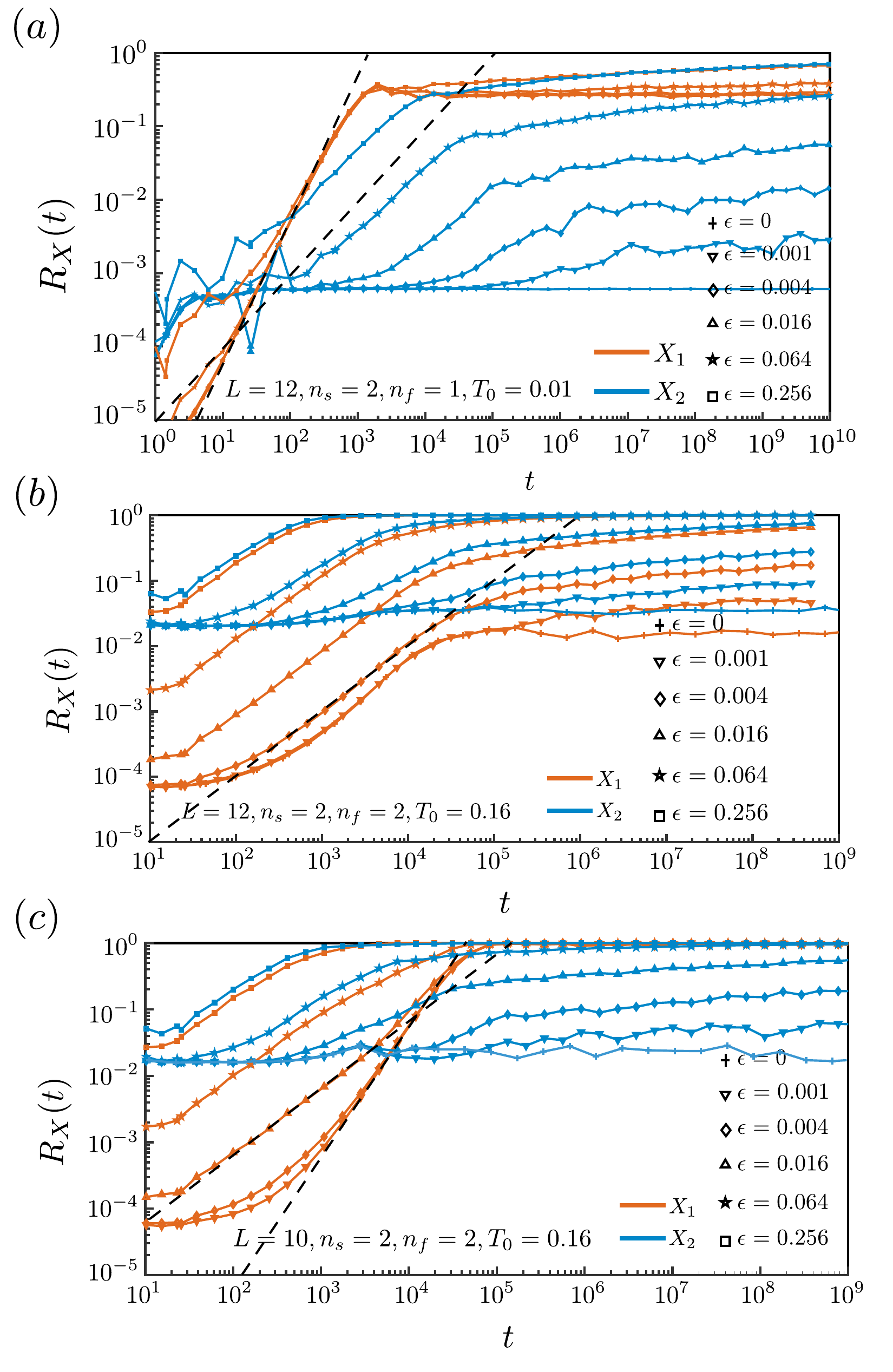}
    \caption{Relaxation dynamics of $X \equiv X_1, X_2$ when the system is driven by the two unitaries $X_1, X_2$, that is, $n_s = 2$. Black dashed lines indicate $\sim t, t^2$ decay. All relevant parameters are indicated in the plots; see the main text for further discussion.}
    \label{fig:ns2_unscaled}
\end{figure}

In Figs.~\ref{fig:ns1_unscaled} and~\ref{fig:ns2_unscaled}, we plot the data for systems driven by the single unitary $X_1$ and the pair of unitaries $X_1,X_2$, respectively. In all cases, the plot corresponds to the relaxation $R_{X} (t)$ of the indicated operator $X$, where we define 

\begin{align}
    R_X (t) = 1 - \abs{\avg{X (t) X (0)}}. 
\end{align}

Here the average is taken over a complete set of states spanning the full Hilbert space. At $t = 0$,  $R_X (t) = 0$ by virtue of the fact that $X^2_i = \mathbb{1}$. At long times, we expect $R_X (t \rightarrow \infty) = 1$ indicating complete relaxation. 

Fig.~\ref{fig:ns1_unscaled} (a) illustrates the broad features alluded to above. For $\epsilon = 0$, we have the ideal polyfractal protocol \cite{agarwal2020dynamical}, which leads to initial rapid relaxation of the operator $X_1$ followed by a plateau which appears to last the course of the simulation. Upon introducing Fibonacci noise, $\epsilon \neq 0$, we see the same plateau at small values of $\epsilon$ eventually giving way to a linear in time ramp which is followed by a final logarithmically slow relaxation regime. 

In Fig.~\ref{fig:ns1_unscaled} (b), we plot more data corresponding to the $n_s = 1, n_f = 1$ case. Different colors indicate different driving periods $T_0$ from $T_0 = 0.01$ where we are driving at a frequency larger than the many-body bandwitdh, to $T_0 = 0.64$, where we are driving at a frequency close to the local energy scales of the system. The fans indicate the span of curves from $\epsilon = 0.001$, that is very minute perturbation to the polyfractal protocol, to $\epsilon = 0.256$, which represents almonst an $\mathcal{O} (1)$ deviation from the ideal protocol. Perhaps the most remarkable aspect of the data is how consistent the behavior remains as the driving period is brought rather close to the local energy scale of the undriven Hamiltonian, that is, $T_0 \sim 1$, and large deviations $\epsilon = 0.256$. In our simulations, we only obtain complete relaxation on $\mathcal{O}(1)$ timescales when the drive period becomes larger than $1$ (not shown).  We also note that, for $\epsilon = 0$, a plateau in the relaxation is obtained on a microscopic time scale $O(1)$, and appears to last for extremely long times; see Fig.~\ref{fig:ns1_unscaled} (c). This corresponds to a prethermal regime wherein memory of a ``rotated" $\tilde X$ \cite{else2017prethermal} persists for an exponentially long time. For $T = 0.01$ this prethmermal regime extends indefinitely in time since the drive frequency exceeds many-body band-width. For finite $\epsilon$, the prethermal plateau gives way to further relaxation which kicks in at a time inversely related to $\epsilon$---in contrast to harmonic high-frequency drives, wherein multiple spins must be flipped to absorb one `photon' from the drive, and thus heating corresponds to a higher order process that kicks in at exponentially late times, the quasiperiodic noise we consider (at finite $\epsilon$) is composed of noise at lower frequencies that can lead to spin flips at earlier times. 

%This is because at finite $\epsilon$, the quasiperiodic drive we consider contains sub-harmonics that can excite the system more rapidly in contrast to periodic high-frequency noise, .

Finally, in Fig.~\ref{fig:ns1_unscaled} (c), we compare relaxation for different numbers of fractal layers, $n_f = 1,2$. It is immediately clear that the main influence of the additional fractal layer is to reduce the initial relaxation determined by the height of the plateau. The plateau is significantly lower since the quasi-conserved operator $\tilde{X}_1$ can be made much closer to the original driving unitary $X_1$ than the simple drive with $n_f = 1$. Due to the fact that the plateau is much lower, the duration of the prethermal plateau is much shorter for $n_f = 2$ and the same $\epsilon$ compared to the $n_f = 1$ case. However, the long time behavior for the $n_f = 1,2$ cases appears is rather similar---we see that an intermediate linear ramp in the $n_f = 2$ case extends in time to the linear ramp in the $n_f = 1$ case [note the orange and blue curves in Fig.~\ref{fig:ns1_unscaled} (c) for the same $\epsilon$]. Thus, the main advantage of driving by multiple fractal layers appears to be in attaining a better approximation of the original driving unitary $X_1$ by  the quasi-conserved operator $\tilde{X}_1$. 

In Figs.~\ref{fig:ns2_unscaled}, we plot data for the relaxation for the case when two unitaries $X_1,X_2$ drive the system. In Fig.~\ref{fig:ns2_unscaled} (a), we see that the second driving unitary $X_2$ appears to show much the same behavior as that of $X_1$ in Fig.~\ref{fig:ns1_unscaled} (a). In particular, we find an initial relaxation on microscopic timescales followed by a plateau which finally gives way to a linear ramp followed by a slow relaxation regime determined by finite $\epsilon$. The behavior of the first driving unitary is more peculiar in that it appears to possess another relaxation mechanism that leads to rapid, approximately $\sim t^2$ relaxation, which is again independent of $\epsilon$, that is, it is intrinsic to the polyfractal protocol. This relaxation then gives way to a plateau which is then followed by similar relaxation behavior as that seen for $X_2$. It is also worth noting that the relaxation due to finite $\epsilon$ in the case of Fibonacci driving appears to affect the relaxation of both $X_1$ and $X_2$ in nearly an identical manner---we see that the relaxation curves for $X_1$ and $X_2$ align at long times wherever the plateau determined by ideal polyfractal driving ($\epsilon = 0$) has been attained and the relaxation due to Fibonacci noise dominates [more evidence of this is seen in Fig.~\ref{fig:ns2_unscaled} (b)]. This suggests that the relaxation of these driving unitaries is indifferent to their specific structure but is largely determined by the noise spectrum associated with Fibonacci noise. 

In Fig.~\ref{fig:ns2_unscaled} (b), we plot results for the $n_s = 2, n_f = 2$ case. Evidently, driving by two fractal layers tends to strongly suppress the $\sim t^2$ ramp observed in the relaxation of $X_1$ in Fig.~\ref{fig:ns2_unscaled} (a). Here we see again the standard behavior---initial relaxation followed by a plateau, which gives way to a linear in time ramp, followed by slow logarithmic relaxation. We again note that the linear in time ramp, which arises only when $\epsilon \neq 0$ appears to affect the relaxation of both $X_1$ and $X_2$ in an almost identical manner as it becomes stronger than the initial relaxation attributed to ideal polyfractal driving. 

Finally, in Fig.~\ref{fig:ns2_unscaled} (c), we plot the same data as computed in Fig.~\ref{fig:ns2_unscaled} (b) but for $L = 10$ instead of $L = 12$. We again observe the presence of the $\sim t^2$ ramp in the relaxation of $X_1$ which is missing in $X_2$. Again, we note that this ramp is intrinsic to the polyfratal protocol itself (the resulting relaxation curve overlays that for $\epsilon = 0.001$ and is not explicitly marked). The comparison between the two figures highlights how this ramp could be related to the commutation relation between the unitaries---they anti-commute for $L = 10$ while commuting for $L = 12$. It is worth noting that when these global unitaries anti-commute, they espouse together a spin-$1/2$ algebra which results in a double degenerate spectrum. In the case they commute, the resulting algebra is Abelian and does not entail a degenerate structure. Such non-Abelian algebras can be more fragile to thermalization~\cite{potter2016symmetry}.  

\subsection{Scaling results}

\begin{figure}[htp]
    \centering
    \includegraphics[width=3.0in]{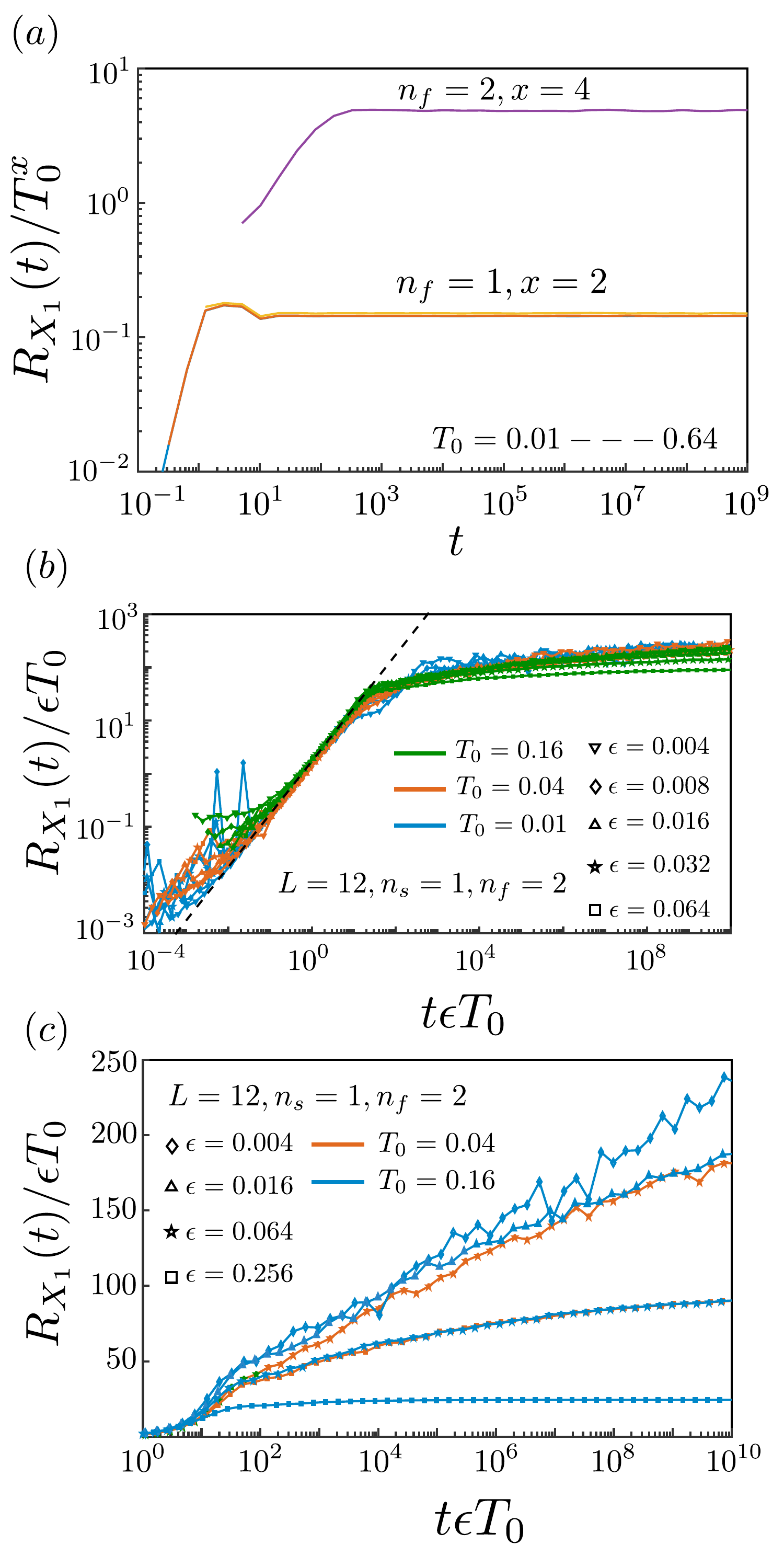}
    \caption{(a) $n_s = 1$, collapse of the plateau at $\epsilon = 0$ for $n_f = 1, x = 2$ and $n_f = 2, x = 4$. Here $T_0 = 0.01 - 0.64$. %\IM{whats x?? also in (a) bad - - -  in $T_0$}
    The collapse is near perfect. (b) Scaling collapse for the intermediate linear ramp. (c) Long-time relaxation dynamics. All relevant parameters are mentioned in the plots; see main text for further discussion.}
    \label{fig:ns1_scaling}
\end{figure}

In Figs.~\ref{fig:ns1_scaling} we attempt to use scaling collapses to extract more information about the general structure of the relaxation curves from the unscaled plots above. We perform these collapses for $n_s = 1$ where it is most feasible given the numerical data accessible. 

In Fig.~\ref{fig:ns1_scaling} (a), we show the scaling collapse for the plateau. The height of the plateau is independent of $\epsilon$. We thus use the ideal polyfractal simulations to obtain this plot. In particular, for $n_f = 1$, we find that the plateau scales as $T_0^2$ in accord with the expectation that the driving kills terms of order $\mathcal{O} \left( T_0 \right)$ that do not commute with $X_1$. Thus, we anticipate a quasi-conserved operator $\tilde{X}_1$ related to $X_1$ by a rotation of order $\mathcal{O} \left( T_0^2 \right)$. For $n_f = 2$, the plateau scales as $T_0^4$ instead of the naively expected $T_0^3$ scaling; this can occur due to symmetry reasons where the absence of certain terms in the physical Hamiltonian naturally eschews $\mathcal{O} (T^3_0)$ terms in the BCH expansion utilized in Eq.~\ref{eq:dyndecoupling} that do not commute with $X_1$. We do not expect this to be the case for more general Hamiltonians than the one we examine. We note that in this plot, we used $T_0 = 0.0$---0.64 and the scaling collapse is near perfect for this range of driving periods.  

In Fig.~\ref{fig:ns1_scaling} (b), we focus on the linear ramp that appears at times after the plateau seen in the polyfractal simulations is obtained. In particular, we normalize the relaxation curve by $\epsilon T_0$ and the time by $1/\epsilon T_0$. The scaling collapse is remarkably good for the large range of values of $\epsilon = 0.004$---$0.064$ and $T_0 = 0.01$---$0.16 $ considered here. This collapse suggests the following results----i) the relaxation rate is constant over the linear ramp and scales $\epsilon^2 T_0^2$, ii) thus, the timescale on which the linear ramp kicks in, $\tau_l$ ,is determined by the time at which relaxation due to this linear heating approaches the height of the plateau in the polyfractcal case, thus, $\epsilon^2 T^2_0 \tau_l \sim T^{n_f}_0$, and iii) the time $\tau_s$ that determines the end of this linear relaxation regime scales as $\tau_s \sim 1/(\epsilon T_0)$. We give qualitative explanation of these results in terms of the Fibonacci noise spectra below. 

In Fig.~\ref{fig:ns1_scaling} (c), we examine the terminal logarithmically slow relaxation regime. Two aspects of the data are evident---i) the relaxation is logarithmic or perhaps even weaker than logarithmic in this regime, and ii) curves for different $T_0$ and $\epsilon$ overlay each other if they have the same product $\epsilon T_0$. This again hints that there are universal features to these long time curves which are largely determined by the single parameter $\epsilon T_0$. We can also explain the presence of this slow logarithmic regime and dependence on the single parameter $\epsilon T_0$ by analyzing the noise spectrum of the Fibonacci noise. 

\subsection{Scaling with system size, random noise, and single spin simulations}

\begin{figure}
    \centering
    \includegraphics[width=3.1in]{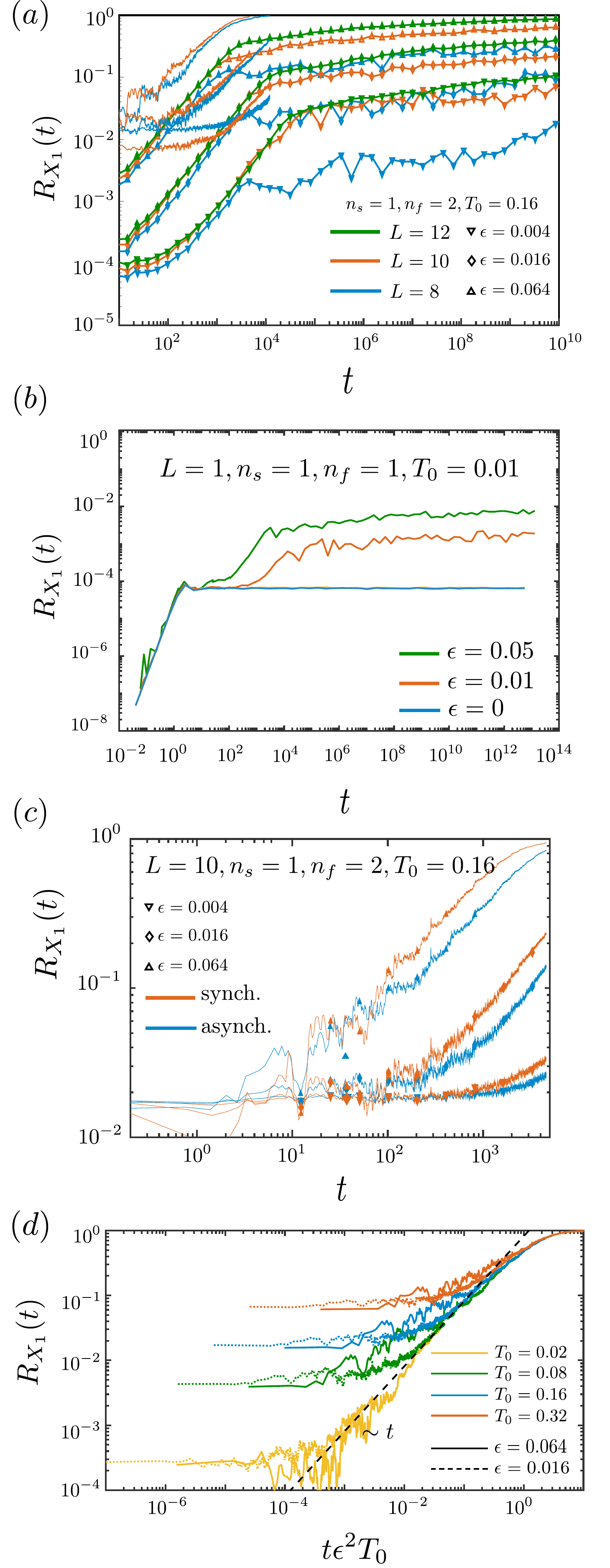}
    \caption{(a) Relaxation curves for different $L$ are compared. Also seen are relaxation under synchronous and asynchronous noise [see also (c)] (b) A single spin driven by Fibonacci noise appears to exhibit similar features to the many-body case. All relevant parameters are stated explicitly in the plots. Note in (a) we do not use markers to distinguish different $\epsilon$ for the synchronous random noise for the sake of clarity---the curves from top to bottom correspond to decreasing values of $\epsilon$ equal to those used for the Fibonacci noise simulations. (d) Scaling collapse for relaxation under random noise in the linear ramp regime. The relaxation rate is given by $\sim \epsilon^2 T_0$ as opposed to $\epsilon^2 T_0^2$ in the Fibonacci case.}
    \label{fig:ns1_scalingL}
\end{figure}

In Fig.~\ref{fig:ns1_scalingL} (a), we focus on the scaling of the data with system size $L$. This is rather challenging to quantify since the system sizes available are rather small, and data for even smaller $L$ appears to be rather jagged. It is natural to assume that the relaxation of the global operator $X_1$ should be proportional to the number of local processes flipping spins, which would suggest a relaxation rate $\sim L$. However, we note that the linear ramp for the $L = 10, 12$ cases appear to lay on top of one another almost perfectly, while there is significant departure between these cuvres and those obtained for $L = 8$. %\IM{at shortest time the curves split consistenly with L. Comment on that?} 
It is thus challenging to draw any conclusions about this given the system sizes accessible numerically. 
%\IM{so, no scaling has been done here at all? should we still call it then scaling in the title of the section?} 
We also plot, in this figure, the relaxation obtained from synchronous random noise. We note that the relaxation curve in this case appears to have no logarithmic relaxation regime, as can be expected due to a lack of long-time correlations in the spectral function of such noise. We indeed find complete relaxation within the simulation times accessible for the largest $\epsilon$. 

%In general, the relaxation rate determining the linear ramp appears to be a factor of $10$ larger than that observed in the Fibonacci noise case; this is likely due to the actual amplitude of the spectral function of the random noise for the same $\epsilon$ being somewhat larger than that of the Fibonacci case; see caption of Fig.~\ref{fig:noisespectrum}. We do not show here, for the sake of clarity, simulations for the asynchronous noise but we note that it is of the same form as the synchronous case, and perhaps surprisingly, has a lower relaxation rate than the latter for all parameters we simulate. 
%\IM{we promised results for the asynchronous noise earlier. We have to show them. I would split off the random stuff from the Fibonacci. 7(a) looks otherwise bit messy}

In Fig.~\ref{fig:ns1_scalingL} (b), we plot the extreme case of $L = 1$ for comparison. Here we show a single spin in a static magnetic field $\vs{B}$, with $\abs{\vs{B}} = 1$, driven by the unitary $X_1 = \sigma_x$. To obtain smooth curves, we average over $1600$ runs corresponding to different values of the spin Hamiltonian governed by $\vs{B}$. We generally see the same features as seen in larger system sizes $L = 8,10,12$---a plateau given by the ideal polyfractal protocol, which gives way to a linear ramp at some late time, followed by the saturation of further relaxation. This again suggests the importance of the noise spectra in determining the full extent of the relaxation curves, to which we now turn. 

Fig.~\ref{fig:ns1_scalingL} (c) shows relaxation under synchronous and asychronous noise for $n_s = 1, n_f = 2$, fixed $T_0 = 0.16$, and various $\epsilon$. In general, we find a linear relaxation regime in this case; In Fig.~\ref{fig:ns1_scalingL} (d), we consider a scaling collapse---the scaling collapse shows that a different relaxation rate $\sim \epsilon^2 T_0$ determines the linear ramp as opposed to the Fibonacci case, where the relaxation rate scales as $\sim \epsilon^2 T_0^2$. We show that these differences can be completely explained by considering the spectral properties of the various noise forms below. 

\subsection{Explanation of relaxation regimes using noise spectra}

We now attempt to explain the observed data and scaling collapses using the noise spectra computed in Sec.~\ref{sec:noisespectrum}. Specifically, we noted there certain features of the Fibonacci noise spectrum which we restate here for better readability---i) the spectrum shows a noise floor which is flat in frequency space which effectively describes the component of the spectrum that appears to mimick random noise over the time $t$. This noise floor scales as $\sim \epsilon^2 T_0^2$ and descreases in weight with time as $\sim 1/t$, ii) the spectrum exhibits spectral peaks associated with the quasiperiodic character of the driving, which also scale as $\sim \epsilon^2 T_0^2$ at $\mathcal{O} \left( 1 \right)$ frequencies and increase in amplitude as $\sim t$, and iii) the peaks have width $1/t$ and show a fractal repetition structure. We also note that in contrast, random synchronous or asynchronous noise exhibit a nearly $t$-independent white noise spectrum that also scales in amplitude as $\epsilon^2 T_0^2$. 

Much of the data presented can be explained by a simple analysis utilizing the above properties of the noise spectra. Since the relaxation curves in the many body case are well reproduced in the single spin simulations [see Fig.~\ref{fig:ns1_scalingL} (b)], it is useful to focus first on the effect of the noise in this simpler setting first. In particular, since we drive with $\sigma_x$, the spin experiences a time-dependent Hamiltonian $B_y (t) \sigma_y + B_z (t) \sigma_z + B_x \sigma_x$. In this case, a Born-Markov analysis can be employed to understand the decay of the autocorrelator $X(t) \equiv \avg{\sigma_x (t) \sigma_x (0)}$. In general, this corresponds to a simple rate equation, with the rate of decay $\Gamma$ of the autocorrelator determined by the amplitude of the noise spectrum at frequency $\omega_x = 2 B_x$~\cite{breuer_theory_2002}. That is, $\Gamma \propto h^2 S(\omega_x)$, where $h$ is the typical amplitude of the magnetic field $B_z, B_y$ that causes relaxation, and $S(\omega_x)$ is the noise spectrum discussed in Sec.~\ref{sec:noisespectrum}. The amplitude of the noise spectrum for the random (white) noise scales as $\epsilon^2 T_0$, while the typical amplitude of noise in the Fibonacci spectrum scales as $\epsilon^2 T_0^2 \omega$, where $\omega = \omega_x \sim h$ is the frequency of interest. Thus, we expect

%In general, for global operators like the $X_i$ considered, relaxation can occur due to both single spin flips, or multi-spin flips. Single spin flips require the system absorbing an energy of the order of the microscopic energy scales from the drive, and via Fermi's Golden rule, the rate of such flips is proportional to the amplitude of the noise spectrum at $\mathcal{O}(1)$ frequencies, times $L$, the number of the spins in the system. Since the noise corresponds to a time-dependent modulation of local terms in the Hamiltonian, multiple spin processes involving $n$ spins occur at order $\mathcal{O} \left( n \right)$ in perturbation theory and are suppressed exponentially in $n$ as $\left( \norm{h}/\omega \right)^n $, where $\omega$ can be approximated as the frequency at which the noise spectrum peaks. The above implies that relaxation of $X_i$ occurs predominantly due to single or few spin processes. 

%When the noise appears uncorrelated on the timescale of the relaxation of the operators $X_i$ under consideration, one can justify a Born-Markov equation of motion for the relaxation of the operator $X_i$ with 

\begin{align}
    \d{\avg{X}}{t} (t) &= - \Gamma \avg{X_i}, \nonumber \\
    \Gamma &\sim h^3 \epsilon^2 T_0^2, \; \; \text{(quasiperiodic)} \nonumber \\
    \Gamma &\sim h^2 \epsilon^2 T_0 \; \; \text{(random)}
    \label{eq:linearramp}
\end{align}

%For the Fibonacci case, note that this rate is computed by \emph{combining} the amplitude from the noise floor in the Fibonacci noise spectra \emph{and} the spectral peaks which encode the quasiperiodicity of the drive. It is the combined spectral weight which remains unchanged at all times even though the noise floor and the height of the peaks change in time. 
%For the Fibonacci case, we 
Note also that all other terms on the right hand side of Eq.~(\ref{eq:linearramp}) corresponding to expectation values of other operators which may otherwise appear in this equation vanish in our case as they rapidly decay on $\mathcal{O}\left(1 \right)$ timescales. 

For $\avg{X}$ close to $1$, or $R_{X} (t) = 1 - \avg{X} \ll 1$, we can linearize and subsequently integrate these equations to obtain the linear relaxation ramp we observe in the data, that is, 

\begin{align}
    R_{X} (t) \approx \Gamma t &\sim h^3 \epsilon^2 T_0^2 t \; \; \text{(quasiperiodic)}, \nonumber \\
    &\sim h^2 \epsilon^2 T_0 t \; \; \text{(random)} .
    \label{eq:linearramprx}
\end{align}

\begin{figure}
    \centering
    \includegraphics[width=3.1in]{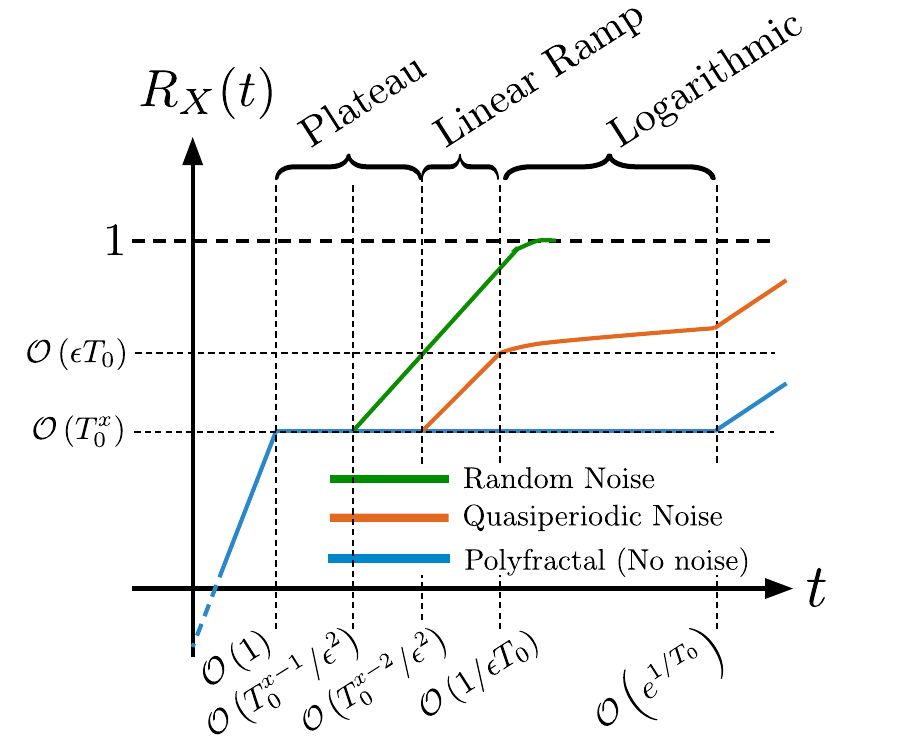}
    \caption{Various relaxation regimes for driving unitaries $X$ are summarized, and the scaling of the relevant timescales is noted. For polyfractal driving without noise, one expects an initial rapid relaxation followed by a prethermal plateau, which gives way to further linear-in-time relaxation beyond an exponentially long time. For quasiperiodic noise, the plateau gives way to a short linear heating regime followed by logarithimically slow relaxation. For truly random noise, the logarithmic relaxation regime is absent, and the operator $X$ relaxes completely in finite time. We anticipate that the logarithmic relaxation observed in the quasiperiodic noise, and the prethermal plateau in the polyfractal case will eventually give way to fast heating beyond the prethermal regime, although our numerical data does not show any sign of such behavior. }
    \label{fig:caricature}
\end{figure}  

This corresponds well with the observed scaling of the initial relaxation of $X$ with $\epsilon, T_0, $ and $t$ for both the quasiperiodic and random noise cases. Since the noise spectrum of the random noise does not change with time, we anticipate that this behavior persists in that case, up to complete relaxation of the spin [although we have to solve Eq.~(\ref{eq:linearramp}) without linearizing it to describe the relaxation curve for times at which $R_X (t) \lesssim 1$.]

Next, we can identify the onset of the slow logairthmic relaxation regime as follows. In the Born-Markov approximation, the spin is susceptible to relax due to noise at a single frequency, $\omega_x = 2B_x$. This rests on the approximation that the bath is incoherent on the timescale of the relaxation of the spin. As the peaks in the quasiperiodic noise spectrum grow sharper, they eventually fail to serve as a bath. This occurs when the relaxation due to the amplitude of these peaks $\sim \epsilon^2 T_0^2 t$ becomes of the order of the width of these peaks $\sim 1/t$ (which is related to the correlation time of the bath), yielding a timescale $\tau_s \sim 1/\epsilon T_0$, as we noted in the scaling collapse in Fig.~\ref{fig:ns1_scaling} (b). In this case, we may think of these peaks as each corresponding to a harmonic drive at the appropriate frequency. When such a peak aligns with the frequency of the spin, $\omega_x$, it can lead to coherent Rabi oscillations for a limited duration which result in rapid oscillations of the autocorrelator but not a decay of memory of the initial state. We see little evidence of such resonant behavior in our simulations. Non-resonant coherent peaks have little effect on the spin. 

Thus, we can expect that beyond the timescale $\tau_s \sim 1/\epsilon T_0$, these coherent peaks can be ignored and do not contribute to further relaxation, leading to the slow relaxation regime, where only the noise floor of the spectrum, which is suggestive of random white noise, contributes to further decay. This naturally yields a relaxation of the form (valid for small $R_{X}$), 

\begin{align}
    \d{R_{X}}{t} &= \Gamma(t) \sim \frac{\epsilon^2 T_0^2}{t}, \; & & \text{or} &
    R_{X} (t) &\sim \epsilon^2 T_0^2 \text{log} (t). 
\end{align}

in the quasiperiodic case. Thus, we see the logarithmically slow relaxation seen in the relaxation curves can emerge from the simple consideration of the noise spectrum of the quasiperiodic noise. This time-dependent effective relaxation rate $\Gamma (t) \sim 1/t$ is of course due to the correlations built into the Fibonacci noise waveform and is absent for truly random noise where we thus do not expect, or see, such a slow relaxation regime. For a single spin, we expect that the above discussion completely captures the relaxation. 

In the many-body case, the operators $X_i$ considered are global operators composed of products of local spins. Relaxation of such operators can occur due to both single spin flips, or multi-spin flips. Single spin flips require the system absorbing an energy of the order of the microscopic energy scales from the drive, and via Fermi's Golden rule, the rate of such flips is proportional to the amplitude of the noise spectrum at $\mathcal{O}(1)$ frequencies. Since the noise corresponds to a time-dependent modulation of local terms in the Hamiltonian, multiple spin processes involving $n$ spins occur at order $\mathcal{O} \left( n \right)$ in perturbation theory and are suppressed exponentially in $n$ as $\left( \norm{h}/\omega \right)^n $, where $\omega$ can be approximated as the frequency at which the noise spectrum peaks. The above implies that relaxation of $X_i$ occurs predominantly due to single or few spin processes. Numerical evidence for this comes from the close resemblance of the relaxation curves of the single spin and many-body cases. However, we do anticipate that eventually, many-spin flip processes which are absent in the single spin system, to lead to linear-in-time relaxation in the many-body case. We do not observe evidence of this in our data, but this can be challenging to discern given the limited system sizes accessible numerically which implies that the many-body level bandwidth is comparable to the bandwidth of single spin flip excitations.  We summarize these findings in a caricature figure, Fig.~\ref{fig:caricature}.

%Finally, in the long time limit, when $R_X (t) \lesssim 1$, we cannot linearlize as in Eq.~(\ref{eq:eq:linearramprx}); solving Eq.~(\ref{eq:linearramp}) without linearizing it leads to a power law decay with an anomalous power law. We observe evidence to this effect in the data where the logarithmic regime appears to correspond to straight lines on a log-log plot but the power law extracted is so small it is virtually impossible to differentiate it from logarithmically slow relaxation. The numerical results and their interpretation in terms of the noise spectra form the major results of this work. 

\section{Conclusions and Discussion}
\label{sec:conclusions}
We now conclude with a summary of our findings and providing context to these results, besides highlighting several noteworthy points. 

Quasiperiodic noise (or driving) has been studied intensively in recent times particularly in the context of driven many-body localized systems~\cite{dumitrescu2018logarithmically,long2021many}, engendering prethermal behavior with multiple time-translation symmetries~\cite{else2020long}, and stabilizing topological edge modes~\cite{peng2018time,friedman2020topological}. Specifically, in these systems, one often finds slow relaxation regimes, and in certain cases, logarithmically slow relaxation has also been observed~\cite{dumitrescu2018logarithmically}. It is anticipated %\IM{anticipated by who?} 
that such a regime arises due to the peculiar spectral properties of quasiperiodic noise.  

In this work, we considered quasiperiodic noise more specifically in the context of the polyfractal protocol~\cite{agarwal2020dynamical}, which shows how one can engineer multiple symmetry generators in an effective Floquet Hamiltonian by driving a many-body quantum system rapidly with unitaries $X_i$ that satisfy $X^2_i = \mathbb{1}$. In particular, we generalized the results of Ref.~\cite{agarwal2020dynamical} by allowing for the unitaries $X_i$ to be applied not at ideally periodic times, but at times that deviate in a qausiperiodic manner from the ideal application times. Crucially, these deviations can be controlled by a small parameter $\epsilon$ which allows us to control systematically the extent of the quasiperiodic noise. The noise we study allows for deviations that are determined by long and short intervals that follow from the Fibonacci recursion relations. This recursive structure allows us to perform numerical simulations up to exponentially long times and glean  numerically, using the parameter $\epsilon$, the effect of the quasiperiodic distortion on the relaxation of the effective symmetry generators $X_i$. 

We found a characteristic relaxation curve that appears in all of our data---an initial relaxation on microscopic timescales to a prethermal plateau determined from the ideal polyfractal or $\epsilon = 0$ case, followed by a linear ramp signifying a constant relaxation rate for a specific long duration that scales as $1/\epsilon T_0$, before eventually giving way to a logarithmically slow relaxation regime. We show that the data can be explained systematically by appealing to the ``noise" spectrum of the Fibonacci drive. In particular, at finite time $t$, the spectrum is composed of two parts---an apparent white noise floor that decreases in amplitude as $\sim 1/t$, and peaks that emerge from the true long-range correlations of the quasiperiodic drive, that increase in amplitude as $\sim t$. We argued that these peaks, along with the noise floor, at initial times contribute together to a constant relaxation rate which scales as $\epsilon^2 T_0^2$, which explains well the scaling collapse of our relaxation curves in the linear ramp regime. Eventually, the peaks emerging from the noise floor grow sharp and cease to contribute to further relaxation. This occurs when their contribution to the relaxation rate exceeds their width, at a time $\tau \sim 1/\epsilon T_0$. This is expected as the quasiperiodic bath then begins to appear coherent and differentiates itself from random white noise. Beyond this, a time-dependent relaxation rate that is derived from the decreasing noise floor which mimics random white noise, and scales in amplitude as $\sim 1/t$, contributes to the relaxation. An effective relaxation rate $\Gamma(t) \sim 1/t$ explains the emergence of the logarithmically slow relaxation in the system. 

Returning to the question of the efficacy of these protocols in engendering effective symmetries, a few comments are in order. First, it is quite remarkable that the prethermal regime in the polyfractal protocol appears to be extremely long lived in the absence of any noise. In particular, the general criteria for such a prethermal regime is that the drive period $T_0$ is much smaller than the inverse local energy scale $\norm{h}$ of the system. We indeed find evidence that such a regime appears only when $T_0 \lesssim \norm{h}$, but a finite duration for the prethermal window, which is known to scale as $\tau \sim e^{\norm{h}/T_0}$ is not evident in our simulations wherein this prethermal window appears to last essentially indefinitely. Furthermore, we observe this prethermal behavior for $T_0 \lesssim 1$ even when driving the system with two unitaries $X_1,X_2$ with the slowest drive period being $16 T_0$ ($n_f = 2$). It appears that despite the slowest driving period being much larger than the inverse local energy scale, the system does not immediately heat up. We anticipate that one explanation for this could be that the effective time-averaged Hamiltonian has a much smaller local energy scale---this is because all terms that do not commute with $X_1, X_2$ are strongly suppressed by the driving. Thus, there is a possible renormalization effect at play whereby, a system with a local energy scale $\approx 1$ can be driven at at frequency $\gtrsim 1$ as to ensure the existence of a prethermal regime, and can then be further driven at a frequency $\gtrsim 1/2$ by another drive as the local energy scale of the effective Floquet Hamiltonian is reduced by a factor of $1/2$ or more due to cancellation of terms that do not commute with the first drive. This requires further investigation. 

Second, our data suggests that the noise considered, both random or quasiperiodic, does not immediately reduce the efficacy of the dynamical decoupling protocols even for fairly large deviations from the perfect application times of $X_i$ (given by large $\epsilon \sim 0.25$), and that its effect can be studied in a linear response framework. This suggests that the polyfractal protocol, and its variants, could be used successfully to implement global and local symmetries in many-body systems that are long lived, provided the noise is not too large. It is, in particular, worth determining how effective such protocols would be in engendering and protecting the coherence of topological edge modes in bosonic spin chains that are experimentally accessible in a variety of quantum simulations. We leave this for future investigation. 

\section{Acknowledgements }

IM was supported by
the Materials Sciences and Engineering Division, Basic Energy Sciences, Office of Science, U.S.
Dept. of Energy.  KA acknowledges funding from NSERC, FRQNT, and a Tomlinson Scholar Award. KA also would like to acknowledge the hospitality of the Aspen Center for Physics where some of the ideas in this manuscript were developed.

\onecolumngrid

\appendix

\section{Fibonacci Driving With Two Unitaries}
\label{app:ns2}

We determine the time-evolution of polyfractal driving under quasi-periodic noise for $n_s = 2$ by driving the system according to the following Fibonacci sequence:
\begin{align}
    U_n & = U_{n-2}^* U_{n-1}, \ \ n \geq 2 \\
    U_1 & = X_1 e^{-i\mathcal{H}T(1-\varepsilon)} = X_1 \mathcal{U}_-\\
    U_2 & = X_1 e^{-i\mathcal{H}T(1+\varepsilon/\phi)} = X_1 \mathcal{U}_+,
\end{align}
where $\varepsilon$, $\phi$ , and the $*$-superscript serve the same purpose as that in the main text. To illustrate by example, consider $n_f = 1$, in which we have the following:
\begin{align}
    U_3 & =\underset{U_1^* \neq U_1}{ \underbrace{X_2 X_1 \mathcal{U}_{-}}} \underset{U_2}{\underbrace{X_1 \mathcal{U}_{+}}} \\
    U_4 & = \underset{U_2^* = U_2}{ \underbrace{X_1 \mathcal{U}_{+}}} \underset{U_3}{\underbrace{X_2 X_1 \mathcal{U}_{-} X_1 \mathcal{U}_{+}}} \\
    U_5 & = \underset{U_3^* \neq U_3}{ \underbrace{X_1 \mathcal{U}_{-}X_2 X_1 \mathcal{U}_{+}}} \underset{U_4}{\underbrace{X_1 \mathcal{U}_{+} X_2 X_1 \mathcal{U}_{-} X_1 \mathcal{U}_{+}}}
\end{align}
In sum, we generally have that $U_{n-2}^* \neq U_{n-2}$. The essence of the issue thereby comes down to finding a recursive scheme for generating the unstarred unitaries as well as their \textit{conjugate} (starred) unitaries. 

\subsection{Recursion relations for 1 fractal layer}

To get a sense of the task at hand, we begin with the $n_f = 1$ case, in which we call the \textit{\textbf{a}ppropriate} unitaries with $X_2$'s placed at even sites \textbf{A} and \textit{\textbf{b}ad} unitaries with $X_2$'s placed at odd sites \textbf{B}. The location of $X_1$'s in $n_f = 1$ is trivial, given that their application is after every period $T$. Note that, when creating $U_n^{\textbf{A}}$, the unstarred unitaries are always in the  \textbf{A} category, whereas the conjugate unitaries are either in the \textbf{A} category or the \textbf{B} category:
\begin{align}
    U_n^{\textbf{A}} = U_{n-2}^{\textbf{A or B}} \cdot U_{n-1}^{\textbf{A}}
\end{align}
The choice of category for the unitary $U_{n-2}^{\textbf{A or B}}$ conjugate to $U_{n-1}^{\textbf{A}}$ is given by determining the expected location $L_{i,j}^{\textbf{A}} (n)$ of its first $X_i$'s applied in the fractal layer $j$.
\begin{figure}
    \centering
    \includegraphics[clip, trim=1cm 16cm 1cm 0cm, width=0.75\textwidth]{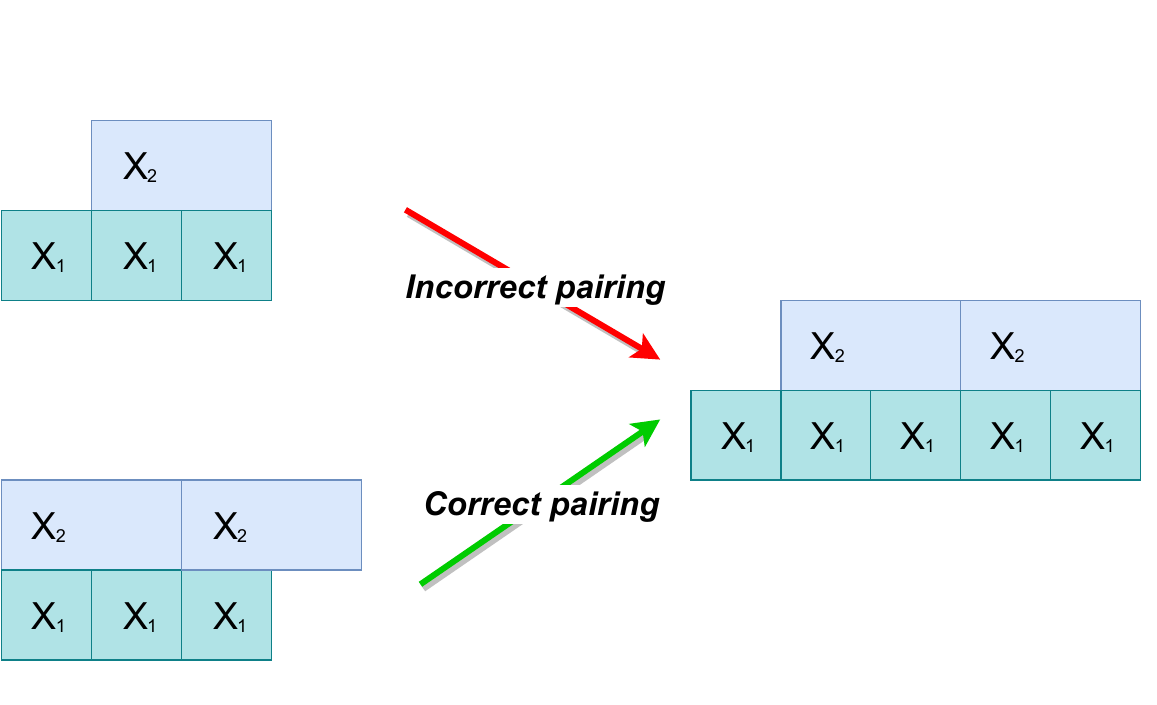}
    \caption{Fractal diagram for $n_f = 1$, in which we see how the $\textbf{A}$ unitary ($t = 5T$) and its conjugate ($t = 3T$) come together to generate the $t = 8T$ $\textbf{A}$ unitary. Note that $X_2$ is placed at an odd side in the conjugate unitary.}
    \label{fig:fractal}
\end{figure}
See figure \ref{fig:fractal} for a visual representation of the following discussion. In other words, $L_{i,j}^{\textbf{A}} (n)$ is simply the fraction of the period of $X_i$, applied in the fractal layer $j$, that remains to be completed in $U_{n-2}^{\textbf{A or B}}$.

The period of a driving unitary $X_i$, applied in fractal layer $j$, is $T = 2^{(j-1)n_s + i - 1} = 2^{2j + i - 3}$ (in units of $T_0$). As a result, the fraction of that period \textbf{already} completed in $U_{n-1}^{\textbf{A}}$ prior to the cut is $F_{n-1} \left( \textup{mod} \ 2^{2j + i - 3} \right)$, where $F_{n-1}$ is the $(n-1)$th Fibonacci number. As a result, the remaining fraction of the period to complete in $U_{n-2}^{\textbf{A or B}}$ is given as follows:
\begin{align}
    L_{i,j}^{\textbf{A}} (n) & = 2^{2j + i - 3} - F_{n-1} \left ( \textup{mod} \ 2^{2j + i - 3} \right ) \\
    L_{1,1}^{\textbf{A}} (n) & = 1 - F_{n-1} \left ( \textup{mod} \ 1 \right ) = 1 \\
    L_{2,1}^{\textbf{A}} (n) & = 2 - F_{n-1} \left ( \textup{mod} \ 2 \right ).
\end{align}
Clearly, $L_{1,1}^{\textbf{A}}(n)$ is trivial. On the other hand, $L_{2,1}^{\textbf{A}} (n)$ indicates that the $X_2$'s are placed at odd sites in the conjugate unitary when $F_{n-1} \left ( \textup{mod} \ 2 \right ) = 1$ and at even sites when $F_{n-1} \left ( \textup{mod} \ 2 \right ) = 0$. As a result, we may find 
\begin{align}
U^{\textbf{A}}_{n} & = \left\{\begin{matrix}
 U^{\textbf{A}}_{n-2} \cdot U^{\textbf{A}}_{n-1} \ \textup{if} \ F_{n-1} \left ( \textup{mod} \ 2 \right ) = 0 \\ 
 U^{\textbf{B}}_{n-2} \cdot U^{\textbf{A}}_{n-1} \ \textup{if} \ F_{n-1} \left ( \textup{mod} \ 2 \right ) = 1
\end{matrix}\right\}
\end{align}
We must now find a means of recursively generating the \textit{bad} unitaries. Similarly to the above discussion, we expect the following:
\begin{align}
    U_n^{\textbf{B}} & = U_{n-2}^{\textbf{A or B}} \cdot U_{n-1}^{\textbf{B}}.
\end{align}
Finding the choice of category for the unitary conjugate to $U_{n-1}^{\textbf{B}}$ comes down to determining the expected location $L_{i,j}^{\textbf{B}}(n)$ of its first $X_i$'s in the fractal layer $j = 1$. At this point, there is an important caveat which we must discuss: given that in $U_{n-1}^{\textbf{B}}$, the $X_2$'s are placed at odd sites, we have that the fraction of its period already completed prior to the cut is $(F_{n-1} \mathbf{-1})\left ( \textup{mod} \ 2^{2j+i-3} \right )$. As a result,
\begin{align}
    L_{1,1}^{\textbf{B}} (n) & = 1 - F_{n-1} \left ( \textup{mod} \ 1 \right ) = 1 \\
    L_{2,1}^{\textbf{B}} (n) & = 2 - \left ( F_{n-1} - 1 \right ) \left ( \textup{mod} \ 2 \right )
\end{align}
As in the previous case, $L_{1,1}^{\textbf{B}}(n)$ is trivial. Furthermore, $L_{2,1}^{\textbf{A}}(n)$ indicates that the $X_2$'s are placed at odd sites in the conjugate unitary when $F_{n-1} \left ( \textup{mod} \ 2 \right ) = 0$ and at even sites when $F_{n-1} \left ( \textup{mod} \ 2 \right ) = 1$:
\begin{align}
U^{\textbf{B}}_{n} & = \left\{\begin{matrix}
 U^{\textbf{B}}_{n-1} \cdot U^{\textbf{B}}_{n-1} \ \textup{if} \ F_{n-1} \left ( \textup{mod} \ 2 \right ) = 0 \\ 
 U^{\textbf{A}}_{n-1} \cdot U^{\textbf{B}}_{n-1} \ \textup{if} \ F_{n-1} \left ( \textup{mod} \ 2 \right ) = 1
\end{matrix}\right\}
\end{align}

The paramount property of this scheme, whose usefulness will be made clear in the $n_f = 2$ case, is that the sequence of either \textbf{A} or \textbf{B} categories repeats: 
\begin{align*}
    U_1^{\textbf{A}} & = X_1 \mathcal{U}_- & U_1^{\textbf{B}} & = X_2 X_1 \mathcal{U}_- \\
    U_2^{\textbf{A}} & = X_1 \mathcal{U}_+ & U_2^{\textbf{B}} & = X_2 X_1 \mathcal{U}_+ \\[-\jot] % you may decrease the vertical space here and in the next line as well.
        & \\
    U_3^{\textbf{A}} & = U_{1}^{\textbf{B}} \cdot U_{2}^{\textbf{A}} & U_3^{\textbf{B}} & = U_{1}^{\textbf{A}} \cdot U_{2}^{\textbf{B}} \\
    U_4^{\textbf{A}} & = U_{2}^{\textbf{A}} \cdot U_{3}^{\textbf{A}} & U_4^{\textbf{B}} & = U_{2}^{\textbf{B}} \cdot U_{3}^{\textbf{B}} \\
    U_5^{\textbf{A}} & = U_{3}^{\textbf{B}} \cdot U_{4}^{\textbf{A}} & U_5^{\textbf{B}} & = U_{3}^{\textbf{A}} \cdot U_{4}^{\textbf{B}} \\[-\jot] % you may decrease the vertical space here and in the next line as well.
        &\\
    U_6^{\textbf{A}} & = U_{4}^{\textbf{B}} \cdot U_{5}^{\textbf{A}} & U_6^{\textbf{B}} & = U_{4}^{\textbf{A}} \cdot U_{5}^{\textbf{B}} \\
    U_7^{\textbf{A}} & = U_{5}^{\textbf{A}} \cdot U_{6}^{\textbf{A}} & U_7^{\textbf{B}} & = U_{5}^{\textbf{B}} \cdot U_{6}^{\textbf{B}} \\
    U_8^{\textbf{A}} & = U_{6}^{\textbf{B}} \cdot U_{7}^{\textbf{A}} & U_8^{\textbf{B}} & = U_{6}^{\textbf{A}} \cdot U_{7}^{\textbf{B}} \\[-\jot] % you may decrease the vertical space here and in the next line as well.
    & \  \ \vdots & & \ \ \vdots \nonumber
\end{align*}
We tabulate the results in \textit{category tables} which display the category of the conjugate (starred) unitaries used creating $U_n^{\textbf{A}}$ (see table \ref{table:1_A}) or $U_n^{\textbf{B}}$ (see table \ref{table:1_B}).
\begin{table}[!ht]
\parbox{.45\linewidth}{
\centering
\begin{tabular}{||c || c | c | c||} 
 \hline
 $n (\textup{mod} \ 3)$ & 2 & 3 & 4 \\
\hline
 $L_{1,1}^{\textbf{A}}(n)$ & 1 & 1 & 1  \\ [0.5ex] 
\hline
 $L_{2,1}^{\textbf{A}}(n)$ & 1 & 2 & 1  \\ [0.5ex] 
 \hline
 $U_{n-2}^{\textbf{A or B}}$ category & B & A & B\\ [0.5ex] 
\hline
\end{tabular}
\caption{Table of categories of the conjugate unitary to $U_{n-1}^{\textbf{A}}$ for $n_s = 2, n_f = 1$}
\label{table:1_A}
}
\hfill
\parbox{.45\linewidth}{
\centering
 \begin{tabular}{|| c || c| c| c||} 
 \hline
 $n(\textup{mod} \ 3)$ & 2 & 3 & 4  \\
\hline
$L_{1,1}^{\textbf{B}}(n)$ & 1 & 1 & 1 \\ [0.5ex] 
\hline
$L_{2,1}^{\textbf{B}}(n)$ & 2 & 1 & 2 \\ [0.5ex] 
 \hline
$U_{n-2}^{\textbf{A or B}}$ category & A & B & A \\ [0.5ex] 
\hline
\end{tabular}
\caption{Table of categories of the conjugate unitary to $U_{n-1}^{\textbf{B}}$ for $n_s = 2, n_f = 1$}
\label{table:1_B}
}
\end{table}

\subsection{Recursion relations for two fractal layers}

The procedure generating the recursion relations for $n_f = 2$ is similar to that of $n_f = 1$, expect that must now keep track of driving unitaries in both $n_f = 1$ and in $n_f = 2$. Consider the sample sequence:
\begin{equation*}
    U_5 = \underset{U_3^{*}}{\underbrace{X_1 \mathcal{U}_- \underline{X_1} X_2 X_1 \mathcal{U}_+}} \underset{U_4}{\underbrace{X_1 \mathcal{U}_+ X_2 X_1 \mathcal{U}_- X_1 \mathcal{U}_+}},
\end{equation*}
where the underlined driving unitary $\underline{X_1}$ indicates that it is applied in $n_f = 2$. In order to appropriately join up $\mathcal{U}_{3}^{*}$ with $\mathcal{U}_{4}$, it must be determined where the sequence of unitaries is terminated in $\mathcal{U}_{4}$. As with the previous case, we have that
\begin{align}
    U_{n}^{\textbf{A}} = U_{n-2}^{\textbf{A or B}} \cdot U_{n-1}^{\textbf{A}},
\end{align}
and as such, we use the general equation derived in the previous section $L_{i,j}^{\textbf{A}} (n) = 2^{2j + i - 3} - F_{n-1} \left ( \textup{mod} \ 2^{2j + i - 3} \right )$. This allows us to tabulate the \textbf{A} category table \ref{table:catA} for $n_s = n_f = 2$. 
\begin{table}[h!]
\centering
 \begin{tabular}{||c | c c c c c c c c c c c c ||} 
 \hline
 $n (\textup{mod} 12)$ & 2 & 3 & 4 & 5 & 6 & 7 & 8 & 9 & 10 & 11 & 12 & 13 \\
 \hline
 $L_{1,1}^\mathbf{A}(n)$ & 1 & 1 & 1 & 1 & 1 & 1 & 1 & 1 & 1 & 1 & 1 & 1 \\ [0.5ex]
 \hline
 $L_{2,1}^\mathbf{A}(n)$ & 1 & 2 & 1 & 1 & 2 & 1 & 1 & 2 & 1 & 1 & 2 & 1 \\ [0.5ex] 
\hline
 $L_{1,2}^\mathbf{A}(n)$ & 3 & 2 & 1 & 3 & 4 & 3 & 3 & 2 & 1 & 3 & 4 & 3\\ [0.5ex]
 \hline
 $L_{2,2}^\mathbf{A}(n)$ & 7 & 6 & 5 & 3 & 8 & 3 & 3 & 6 & 1 & 7 & 8 & 7\\ [1ex] 
 \hline
 \hline
 $U_{n-2}^{*}$ category & B & C & D & E & A & E & E & C & F & B & A & B \\
 \hline
\end{tabular}
\caption{Table of categories of the conjugate unitary to $U_{n-1}^{\textbf{A}}$ for $n_s = 2, n_f = 2$}
\label{table:catA}
\end{table}
As with the $n_f = 1$ case, one may cycle through the category table, which helps speed up numerical simulations instead of needing to compute the remainder of large Fibonacci numbers at every iteration. However, the number of categories, and consequently the complexity of the recursion relations, increases exponentially with $n_s$ and $n_f$. Specifically, the cycle length is given as $\ell = 3 \cdot 2^{n_s n_f - 2}$, whereas the number of categories may be computed thusly $K = 2^{n_s n_f - 1}$.

As with the $n_s = 2, n_f = 1$ case, we need also find the recursion relations allowing us to generate the wealth of unitaries other than those in the $\mathbf{A}$ category. To do so, we adjust the general $L_{i,j}$ equation. Specifically, the equation becomes the following for a given category \textbf{cat}:
\begin{align}
    L_{i,j}^{\textbf{cat}} (n) = 2^{2j + i - 3} - ( F_{n-1} - k_{i,j}^\textbf{cat}) \left ( \textup{mod} \ 2^{2j + i - 3} \right ),
\end{align}
where $k_{i,j}^{\textbf{cat}}$ is the location of the first appearance of $X_i$ in $n_f = j$ in unitaries belonging to the given category. For instance, we have that $\vec{\textbf{k}}^{\textbf{B}} = (k_{1,1}^{\textbf{B}},k_{2,1}^{\textbf{B}},k_{1,2}^{\textbf{B}},k_{2,2}^{\textbf{B}}) = (1,1,3,7)$ and, naturally, $\vec{\textbf{k}}^{\textbf{A}} = (k_{1,1}^{\textbf{A}},k_{2,1}^{\textbf{A}},k_{1,2}^{\textbf{A}},k_{2,2}^{\textbf{A}}) = (0,0,0,0)$. This allows us to generate all the necessary category tables. This framework may be easily extended to generate the recursion relation for all combinations of $n_s,n_f$.

\section{Fibonacci Driving With A Single Unitary}
\label{app:ns1}

With the generalized protocol established in the previous appendix, it is easy to determine the recursion relations for Fibonacci driving with a single unitary $X_1$. The location equation yields the following:
\begin{align}
    L_{1,j}^{\textbf{A}} (n) = 2^{2(j-1)} - F_{n-1} \left ( \textup{mod} \ 2^{2(j-1)} \right ),
\end{align}
which gives the position of the first application of $X_1$ in the fractal layer $n_f = j$. If $n_f = 1$, we have that $L_{1,1}^{\textbf{A}} (n) \equiv 0$ trivially. For $n_f = 2$, we have that $L_{1,2}^{\textbf{A}} (n) = 4 - F_{n-1} \mod \ 4$. Using this, we determine the category table \ref{table:ns1_A}, which bears the same recursion complexity as the $n_s = 2, n_f = 1$ case, which is expected given that both $\ell$ and $K$ depend on the product $n_s n_f$. Furthermore, using the generalized location equation, we may also determine the \textbf{B} category table \ref{table:ns1_B}, which also exhibits similarity to the $n_s = 2, n_f = 1$ case.
\begin{table}[!ht]
\parbox{.45\linewidth}{
\centering
 \begin{tabular}{|| c || c| c| c||} 
 \hline
 $n(\textup{mod} \ 3)$ & 2 & 3 & 4  \\
\hline
$L_{1,1}^{\textbf{A}}(n)$ & 1 & 1 & 1 \\ [0.5ex] 
\hline
$L_{1,2}^{\textbf{A}}(n)$ & 2 & 3 & 2 \\ [0.5ex] 
 \hline
$U_{n-2}^{\textbf{A or B}}$ category & B & A & B \\ [0.5ex] 
\hline
\end{tabular}
\caption{Table of categories of the conjugate unitary to $U_{n-1}^{\textbf{A}}$ for $n_s = 1, n_f = 2$}
\label{table:ns1_A}
}
\hfill
\parbox{.45\linewidth}{
\centering
 \begin{tabular}{|| c || c| c| c||} 
 \hline
 $n(\textup{mod} \ 3)$ & 2 & 3 & 4  \\
\hline
$L_{1,1}^{\textbf{B}}(n)$ & 1 & 1 & 1 \\ [0.5ex] 
\hline
$L_{1,2}^{\textbf{B}}(n)$ & 3 & 2 & 3 \\ [0.5ex] 
 \hline
$U_{n-2}^{\textbf{A or B}}$ category & A & B & A \\ [0.5ex] 
\hline
\end{tabular}
\caption{Table of categories of the conjugate unitary to $U_{n-1}^{\textbf{B}}$ for $n_s = 1, n_f = 2$}
\label{table:ns1_B}
}
\end{table}
Cases with larger $n_f$ may be derived with the generalized location equation in fashions similar to those presented.

\bibliography{fibonacci}

\end{document}